\documentclass[pra,twocolumn,showpacs,letterpaper,showpacs,superscriptaddress,nofootinbib]{revtex4-1}
\usepackage{graphicx,amsmath,amssymb,amsfonts,latexsym,color,dcolumn,bm,subfigure}
\usepackage{mathrsfs}
\usepackage{capt-of,multirow}
\usepackage{float}

\newcommand{\ba}{\begin{align}}
\newcommand{\eA}{\end{align}}

\DeclareMathOperator{\sech}{sech}
\def\mymathhyphen{{\hbox{-}}}
\newcommand{\gz}{g_{{\bf q},0j}}
\newcommand{\g}{g_{{\bf q},j}}
\newcommand{\gzn}{g_{{\bf q},0}}

\begin{document}

\title{Dimensional transformation of defect-induced noise, dissipation, and nonlinearity}

\author{R. O. Behunin}
\affiliation{Department of Applied Physics, Yale University, New Haven, Connecticut 06511, USA}
\author{F. Intravaia}
\affiliation{Max-Born-Institut, 12489 Berlin, Germany}
\author{P. T. Rakich}
\affiliation{Department of Applied Physics, Yale University, New Haven, Connecticut 06511, USA}

\date{ \today}

\begin{abstract}
In recent years, material-induced noise arising from defects has emerged as an impediment to quantum-limited measurement in systems ranging from microwave qubits to gravity wave interferometers. As experimental systems push to ever smaller dimensions, extrinsic system properties can affect its internal material dynamics. In this paper, we identify surprising new regimes of material physics (defect-phonon and defect-defect dynamics) that are produced by dimensional confinement. Our models show that a range of tell-tale signatures, encoded in the characteristics of defect-induced noise, dissipation, and nonlinearity, are profoundly altered by geometry. Building on this insight, we demonstrate that the magnitude and character of this material-induced noise is transformed in microscale systems, providing an opportunity to improve the fidelity of quantum measurements. Moreover, we show that many emerging nano-electromechanical, cavity optomechanical and superconducting resonator systems are poised to probe these new regimes of dynamics, in both high and low field limits, providing a new way to explore the fundamental tenets of glass physics.
\end{abstract}


\maketitle

\section{Introduction}
The demand for ever higher-fidelity quantum measurement and information processing using photonic, microwave, and phononic excitations has invigorated interest in the quantum origins of noise and decoherence within materials. As scientists and engineers seek to reduce the noise in their quantum systems, they have been operating at lower and lower temperatures. This strategy minimizes noise induced by thermal fluctuations, but low temperatures also reveal a fundamental problem: excess dissipation is produced by low-energy defect centers \cite{Anderson72,Phillips72,Jackle72,Phillips87}. These defect centers, termed two-level tunneling-state systems (TLSs), behave much like atoms, and couple strongly to electromagnetic (EM) and phononic fields; the adverse impact of these defects becomes more acute as temperatures are reduced. While TLS defects are inherent to amorphous materials \cite{Phillips87,Pohl02}, they occur and can be induced in crystalline media in a number of ways, raising many questions about their appearance in a range of new quantum systems \cite{Simmonds04,Astafiev04,Martinis05,Zolfagharkhani05,Gao07,O'Connell08,Regal08,Neeley08,Anetsberger08,Arcizet09,Park09,Lindstrom09,Wang09,Shalibo10,Macha10,Palomaki10,Lisenfeld10a,Lisenfeld10,Hoehne10,Riviere11,Pappas11,Sage11,Galliou11,Khalil11,Grabovskij12,Faoro12,Goryachev12,Galliou13,Goryachev13,Goryachev13c,Faust14,Tao14,Burnett14,Meenehan14,Skacel15,Faoro15} (e.g. Fig. 1).
\begin{figure}[b]
\begin{center}
\includegraphics[width=2.5in]{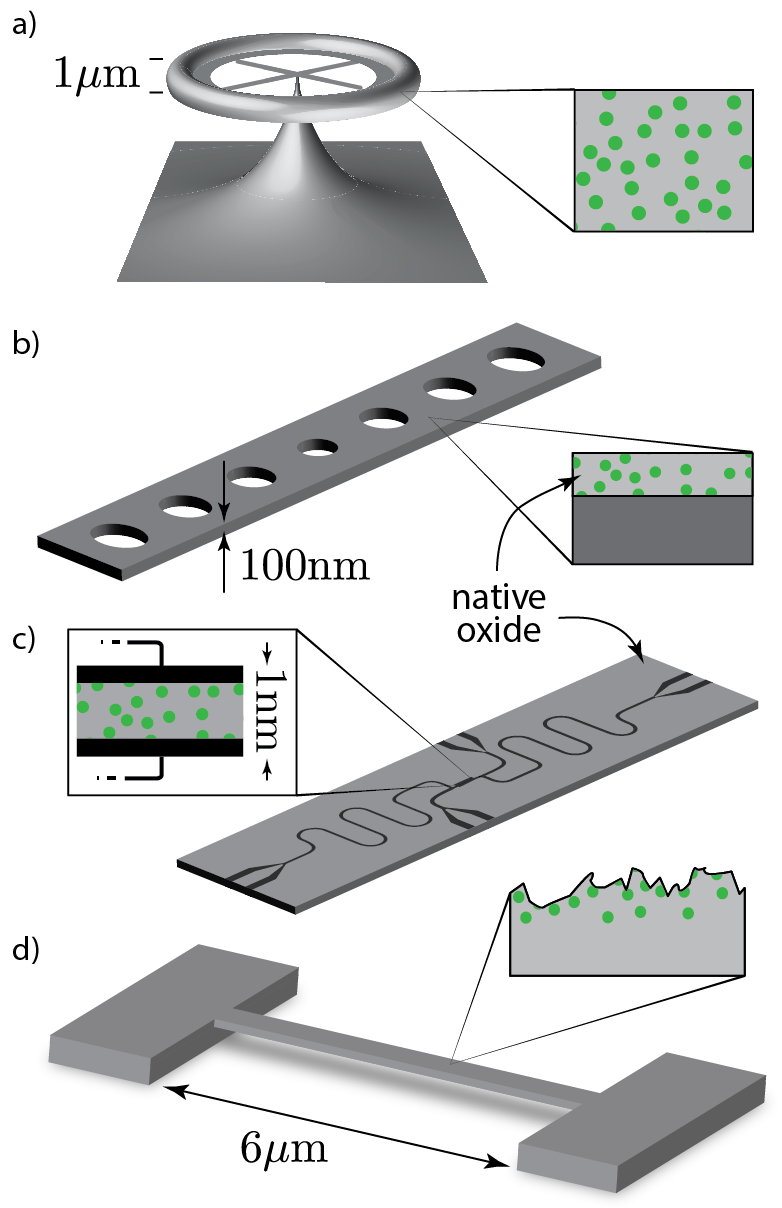} 
\caption{Systems where defect-phonon and/or defect-photon interactions may be dimensionally reduced: a) silica microtoroids, b) silicon optomechanical systems, c) superconducting qubits, and d) nanoelectromechanical systems}
\label{IntroFigure}
\end{center}
\end{figure}

Over the past several decades the properties of dense ensembles of low-energy defects have been extensively studied in bulk amorphous materials \cite{Phillips87,Pohl02}. TLS-based models of defect physics accurately capture (predict) the observed phononic, optical and microwave loss and noise characteristics. To date, it has been appropriate to use the bulk models of TLS ensembles in the vast majority of systems. However, with the emergence of high-confinement phononic, nanoelectromechanical (NEMS), optomechanical, and microwave systems, circumstances arise where it is no longer appropriate to invoke the many properties of bulk systems \cite{Seoanez07a,Seoanez07b,Fu89,Hoehne10,Sage11,Ramos13} (see Fig. \ref{IntroFigure}). As phonons and EM waves are confined to ever decreasing volumes, it is unclear how dissipation, noise, and nonlinearity will be altered \cite{Fu89}. 

In this paper we show that the magnitude and character of noise, dissipation, and nonlinearity arising from low-energy defects is radically modified at mesoscales, offering new challenges and opportunities for emerging NEMS, optomechanical, and superconducting resonator technologies (see Fig. \ref{IntroFigure}). 
In contrast with bulk systems, TLSs in mesoscale structures interact with a zoo of hybridized excitations where small mode volumes, dispersion, and changes of the density of states strongly enhance (or suppress) coupling to the EM and phononic fields. Consequently, appreciable modulation of TLS-induced noise, dissipation, and nonlinearity can be achieved by scaling system dimensions \cite{Hoehne10,Sage11}, opening a path toward extended coherence times in qubits \cite{Sage11}, high-fidelity quantum information processing, phonon lasing \cite{Vahala08,Grudinin10}, and new regimes of quantum nonlinearity \cite{Ramos13}.  

Starting from the established phenomenological model of low-energy defects \cite{Phillips87}, we use techniques of quantum statistical mechanics to characterize the dynamics of individual TLSs in mesoscale structures. These defects interact with confined EM and phononic fields as well as other defects. We find that the excited state lifetime for defects is geometrically reshaped in waveguides and resonators, for example, by Purcell enhancement \cite{Purcell46,Kleppner81} of emission into slow group velocity or stationary resonator modes. In addition, we show that key defect dephasing processes depend sensitively on the system dimension, being thoroughly suppressed in 1$D$ waveguides (see text for details). Utilizing the intimate connection between defect dynamics, noise, and absorption we demonstrate that large geometric modulation and reshaping of TLS-induced noise spectra, huge system-dependent changes in the saturation characteristics of dissipation, and unprecedented reduction in the fundamental TLS-limited dissipation floor is achievable with skillful device design.

The paper is organized as follows. In Sec. II we discuss the variety of systems affected by TLS-induced losses. In Sec. III we lay the foundations for the theory of defect-induced noise applicable to reduced dimensional systems. In Sec IV. defect decay and dephasing rates are discussed, the acoustic analog of Purcell enhancement is derived and the phenomenology of spectral diffusion is developed. Results for defect-induced dissipation are presented.  Dissipation in $D$-dimensional systems, saturation of resonant absorption, and the effects of a gapped phononic spectrum are discussed. The conditions for dimensional reduction are discussed topic by topic. In conclusion, general results and insights of the reduced dimensional theory are codified.

\section{Material induced noise and dissipation} 
\begin{figure}[h]
\begin{center}
\includegraphics[width=0.45\textwidth]{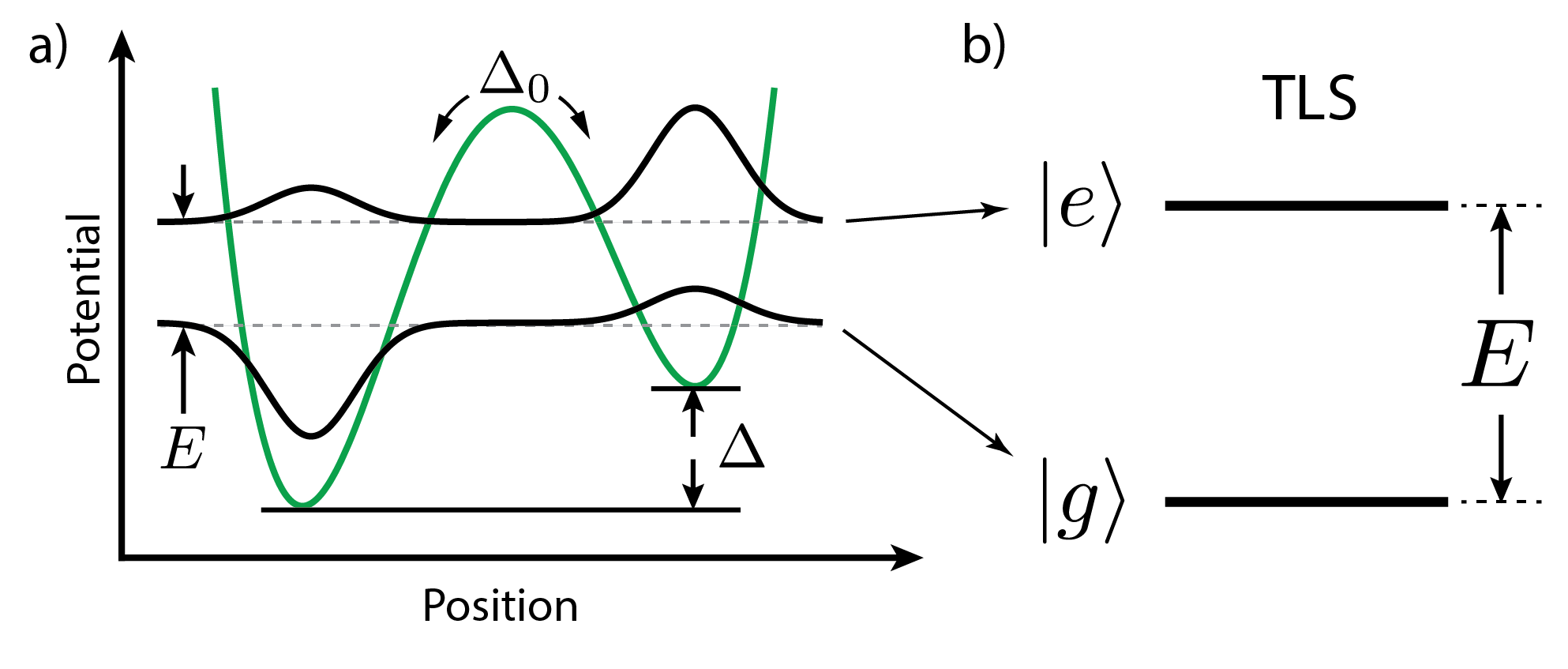}
\caption{a) Double-well potential of asymmetry $\Delta$ for a tunneling state defect. b) Excited $\left| e \right>$ and ground $\left| g \right>$ eigenstates are gapped by energy $E$.}
\label{TLS-diagram}
\end{center}
\end{figure}

The tunneling state model (TSM) of low-energy defects was introduced in the 1970s to explain the anomalous low-temperature properties of the heat capacity of glasses \cite{Zeller71,Anderson72,Phillips72,Jackle72,Phillips87}. Since the inception of the TSM, exhaustive studies have established the TLS concept as a cornerstone of glass physics as it provides a tractable model to describe the low-temperature characteristics of highly disordered media \cite{Phillips87}. Tunneling states are hypothesized to arise from atoms (or groups of atoms) residing in asymmetric double-well potentials that are believed to be inherent to amorphous materials (see Fig. \ref{TLS-diagram}). In recent years, however, it has been realized that TLSs are ubiquitous: they are induced in crystalline materials by irradiation \cite{Laermans79}, impurities \cite{Phillips87,Narayanamurti70,Kleiman87}, dislocations \cite{Hoehne10}, and oxidization \cite{Simmonds04,Martinis05,Gao07,O'Connell08,Wang09,Palomaki10,Sage11,Khalil11}, and they appear at surfaces and interfaces \cite{Zolfagharkhani05,Gao07,Lindstrom09,Wang09,Macha10,Pappas11,Sage11,Khalil11,Burnett14}, making them an important consideration for noise and dissipation in a variety of quantum systems.  
 
Numerous studies have shown that TLSs interact strongly with phonons \cite{Golding73,Golding76,Golding76b}, light, and microwaves \cite{Phillips87}, allowing them to absorb and emit EM and acoustic energy. While an individual TLS typically couples to EM and acoustic modes at comparable rates, the relaxation of TLSs is dominated by phonon emission.  This is because the large disparity between the light and sound speeds makes the phonon density of states (and emission rates) typically orders of magnitude larger. Hence, through absorption TLSs act to dissipate coherent excitations, and through stochastically driven emission (dominated by phonons) TLSs act as a source of EM and acoustic noise, these phenomena being intimately connected through the fluctuation-dissipation relation. 

TLS-induced noise (dissipation) has two signatures: enhancement at low temperature, and high-power suppression (saturation). These signatures have been observed in an increasing number of mesoscopic and macroscopic systems seeking to utilize quantum coherence for both information processing and metrology. For instance, TLSs reside in tunnel junctions \cite{Simmonds04,Martinis05,Shalibo10,Lisenfeld10,Lisenfeld10a,Palomaki10,Grabovskij12}, oxide surface layers \cite{Gao07,O'Connell08,Wang09,Sage11,Khalil11,Skacel15}, and at interfaces of superconducting circuits \cite{Lindstrom09,Wang09,Burnett14} and electro-optomechanical systems \cite{Regal08} (Fig. 1); similarly, they are found in crystalline bulk acoustic wave resonators at dislocations and impurities \cite{Galliou11,Goryachev12,Galliou13,Goryachev13,Goryachev13c}. Tunneling-states are endemic to amorphous systems where they limit the quality factor of optomechanical microtoroids \cite{Arcizet09,Anetsberger08,Riviere11} and NEMS and micro-electromechanical systems (MEMS)  \cite{Zolfagharkhani05,Hoehne10,Faust14,Tao14} (see Fig. 1), and they lead to Brownian motion of mirror coatings that degrades the finesse of interferometers used for gravity wave astronomy \cite{Martin14,Hamdan14}. Hence, the mastery of defect-physics is essential to the manipulation of noise and dissipation in mesoscopic systems, and provides an avenue toward radical improvements in the performance of cutting-edge technologies.   

As an ancillary outcome, the exploration of defect physics in mesoscale systems directly probes the foundations of glass physics \cite{Fu89}. Despite the success of the tunneling state concept, the microscopic nature and origin of TLSs is still unclear \cite{Phillips87}, and in addition, the TSM can only explain the apparent universality of many low temperature glass properties \cite{Leggett91,Pohl02} as resulting from a fortunate fine tuning. These puzzles have inspired researchers to look for alternative theories to the TSM \cite{Yu88,Yu89,Leggett91,Vural11,Leggett13} that are testable in reduced-dimensional systems \cite{Fu89}.  

\section{Theory of defect-phonon/defect-photon interactions in mesoscopic systems}
\begin{figure}[h]
\begin{center}
\includegraphics[width=0.45\textwidth]{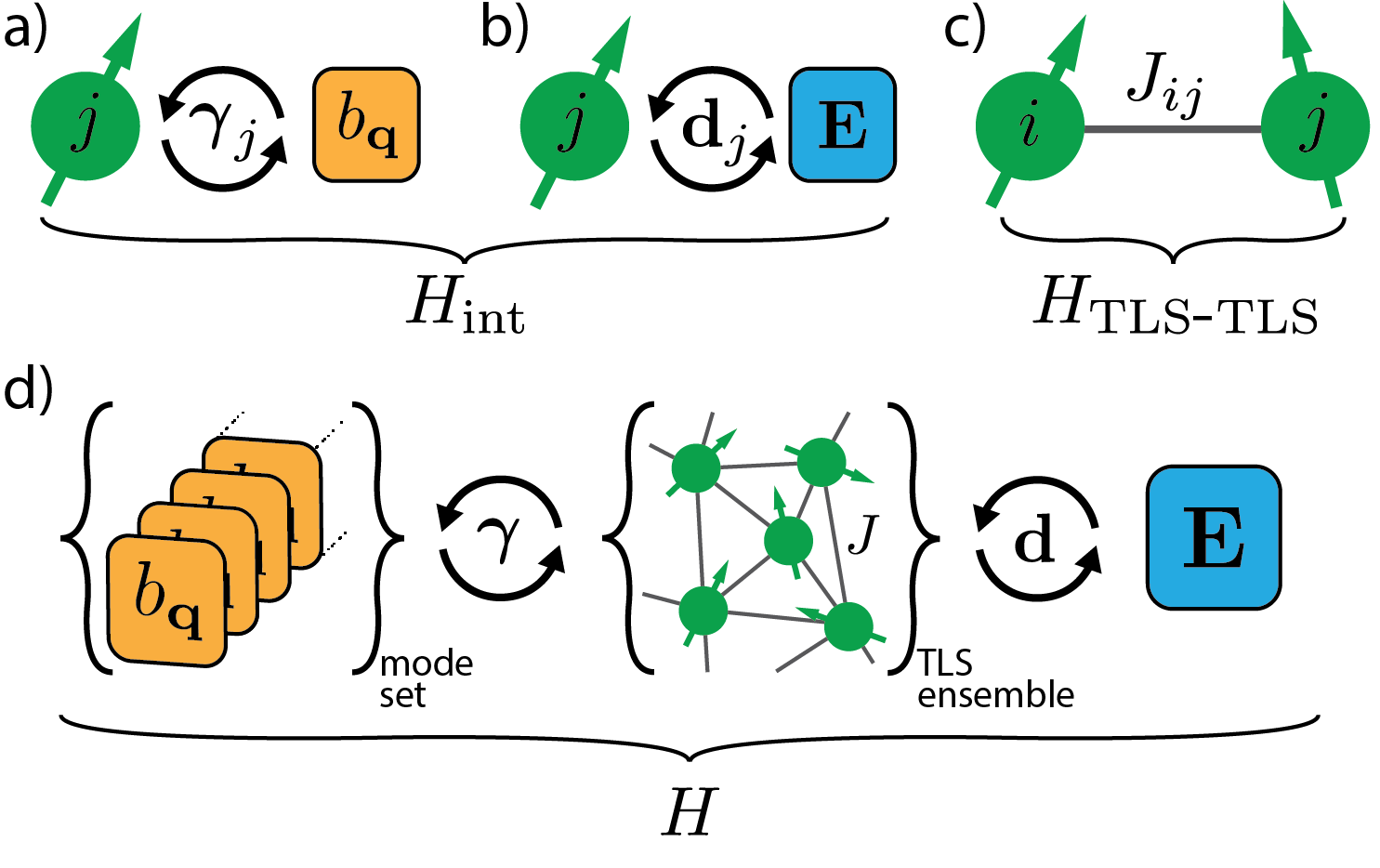}
\caption{Illustration of system Hamiltonian: a) interaction of $j$th defect with ${\bf q}th$ phonon mode, b) interaction of $j$th defect with the EM field, c) defect-defect interactions, and d) illustration of total coupled system.}
\label{HamiltonianPic}
\end{center}
\end{figure}

At low temperatures tunneling state defects can be modeled as effective two-level systems with a spin analogy \cite{Anderson72,Jackle72,Phillips87}. 
The total Hamiltonian $H$ describing the interaction of TLSs with phonons is given by 
\begin{align}
\label{Hamiltonian}
& H = \sum_{\bf q} \hbar \Omega_{\bf q} b^\dag_{\bf q} b_{\bf q} +  \sum_{j} \frac{1}{2} E_j \sigma_{z,j}  + H_{\rm int} + H_{\rm TLS\mymathhyphen TLS}.
\end{align}
The first term on the right-hand side is the free Hamiltonian for the confined phonon field where ${\bf q}$ is a collective phonon mode index, and $b_{\bf q}$, $b^\dag_{\bf q}$, and $\Omega_{\bf q}$ are the respective annihilation operator, creation operator, and angular frequency of the ${\bf q}$th phonon mode.  The second term is the Hamiltonian for the ensemble of non-interacting TLSs. The sum on $j$ adds the contribution to the system energy from each defect with respective energy $E_j = \sqrt{\Delta^2_j+\Delta_{0j}^2}$, double-well asymmetry $\Delta_j$, tunneling strength $\Delta_{0j}$, and where $\sigma_{k,j}$ is the $k$th-component of the defect's Pauli spin operator \cite{Phillips87}. The final two terms, $H_{\rm int}$ and $H_{\rm TLS\mymathhyphen TLS}
$, describe the interactions of the coupled system, schematically shown in Fig. \ref{HamiltonianPic} and described below. In addition, TLSs interact with the confined electromagnetic field, but the Hamiltonian for photons is not included above for compactness.

Perturbations of the double-well potential asymmetry $\Delta$ by elastic strain couple defects and phonons, and EM coupling is produced by charge transfer connected with TLS state transitions. These two coupling mechanisms are described by the interaction Hamiltonian given by  
\begin{align}
\label{Interaction-Hamiltonian}
 H_{\rm int} =  \sum_{j} & \left(\frac{\Delta_{0j}}{E_j} \sigma_{x,j} + \frac{\Delta_j}{E_j} \sigma_{z,j} \right) \nonumber \\
 & \quad \quad \times  [{\bf d}_j \cdot {\bf E}({\bf r}_j)+ \boldsymbol{\gamma}_j: \underline{\xi}({\bf r}_j) ]
\end{align}
where $\boldsymbol{\gamma}_j$ is the $j$th defect's deformation potential tensor, quantifying an energy shift induced by strain, and ${\bf d}_j$ quantifies the defect's electric dipole coupling \cite{Phillips87}. The respective electric field and strain tensor evaluated at the position of the defect are denoted by
${\bf E}({\bf r}_j)$ and $\underline{\xi}({\bf r}_j)$, and the shorthand ${ \boldsymbol{\gamma}_j: \underline{\xi}}$ is defined by  $\gamma_j^{ab} \xi_{ab}$ where the Einstein summation convention is used. The strain is defined as $\xi_{ab} \equiv (\partial_a u_b + \partial_b u_a)/2$ where $\partial_a$ is a spatial derivative in the $a$th direction, and $u_b$ is the $b$th component of the elastic displacement field. 

\begin{figure}[h]
\begin{center}
\includegraphics[width=0.45\textwidth]{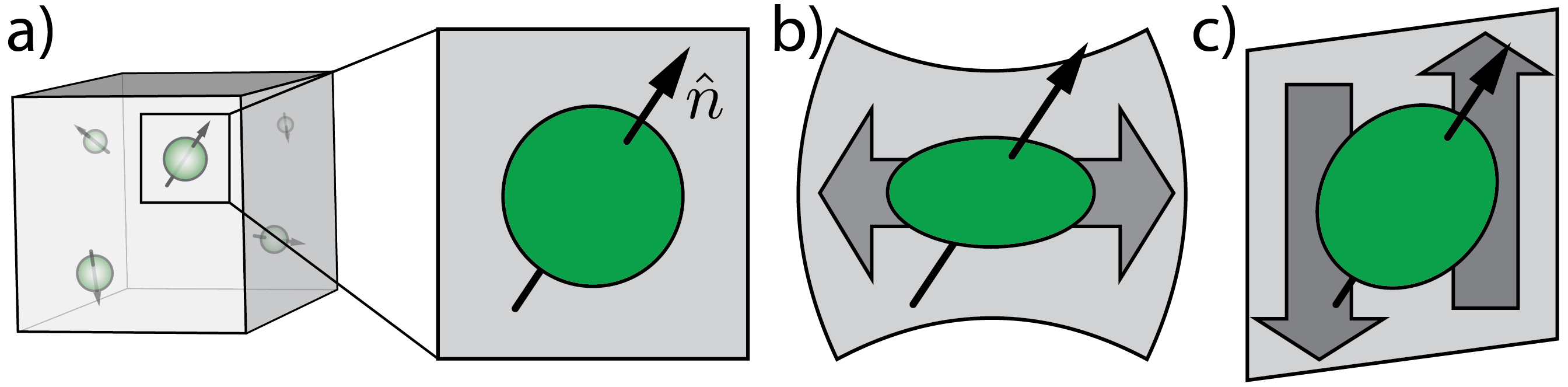}
\caption{Illustration of defect-strain coupling mechanisms. a) Arbitrarily oriented defect in an undeformed elastic body. b) Defect in elastic body undergoing compressional motion, induced defect elastic dipole proportional to $\gamma_\ell(\hat{n})$. c) Defect in an elastic body undergoing shear motion, induced defect elastic dipole proportional to $\gamma_t(\hat{n})$ (see Eqs. \ref{DefPots}).}
\label{defectCoupling}
\end{center}
\end{figure}

The tensor structure of the deformation potential can be worked out from the orientation of a defect's dipole moment and from the symmetry properties of the system material \cite{Anghel07}. For amorphous media, the contraction of the deformation potential of an arbitrarily oriented TLS and the strain tensor is given by
${ \boldsymbol{\gamma}: \underline{\xi}
= \tilde{\gamma} [(1-2\zeta) {\rm tr} \underline{\xi} + 2 \zeta \hat{n} \cdot \underline{\xi} \cdot \hat{n} ] }
$ where tr denotes trace, and $\hat{n}$ is a unit vector parallel to the TLS elastic dipole moment, i.e. $\hat{n} = \sin \theta (\cos \phi \ \hat{x} +  \sin \phi \ \hat{y}) + \cos \theta \ \hat{z} $ where $\phi$ and $\theta$ are the azimuthal and polar angles in spherical coordinates \cite{Anghel07}. The deformation potential for longitudinal and transverse waves, averaged over all defect orientations, is given respectively by $\gamma_\ell^2 = \tilde{\gamma}^2(15 - 40 \zeta + 32 \zeta^2)/15$ and $\gamma^2_t = 4\tilde{\gamma}^2 \zeta^2/15$ (see Fig. \ref{defectCoupling}) \cite{Anghel07}. For silica $\gamma_\ell \approx 1$ eV, and $\gamma_t \approx \gamma_\ell/\sqrt{2}$ \cite{Golding73,Golding76,Golding76b,Graebner86} which results in two possible values for the parameter $\zeta = \{0.57,1.10\}$, of which we use $\zeta = 0.57$. It is unknown whether such a form for the deformation potential will apply to thin films or microwires where the bulk system symmetries may be broken. 

The defect dipole electric coupling has been measured in dielectric absorption experiments and takes the value $|{\bf d}| = 0.3-3.3$ Debye for silica \cite{Bernard78}. Hence, for comparable mode volumes photons and phonons couple to defects with the nearly same strength.  

The geometry of a system is encoded in the mode structure and dispersion of the EM and acoustic fields. These effects are elicited by representing the elastic field as a mode summation ${\bf u}({\bf x}) = \sum_{\bf q} \sqrt{\frac{\hbar}{2 \Omega_{\bf q}}} {\bf u}_{\bf q}({\bf x}) b_{\bf q} + {\rm H.c.}$ where ${\bf u}_{\bf q}$ is the spatial eigenfunction of the ${\bf q}$th mode. ${\bf u}_{\bf q}({\bf x})$ satisfies the time-harmonic Christoffel equation \cite{Rose99} and the orthonormality relation ${\int d^3 x \rho({\bf x}) {\bf u}_{\bf q}^*({\bf x}) {\bf u}_{\bf q'}({\bf x}) = \delta_{\bf q q'}}$ where $\rho({\bf x})$ is the material density. For an isotropic material
${\mu \nabla^2 {\bf u}_{\bf q}({\bf x}) + (\lambda + \mu) \nabla \nabla \cdot {\bf u}_{\bf q}({\bf x}) = - \rho \Omega^2_{\bf q} {\bf u}_{\bf q}}({\bf x})$ where $\lambda$ and $\mu$ are the, generally space-dependent, Lam\'e coefficients of the system \cite{Rose99}. Likewise, the strain field can be expressed as  
$\underline{\xi} = \sum_{\bf q} \sqrt{\frac{\hbar}{2 \Omega_{\bf q}}} \underline{\xi}_{\bf q} b_{\bf q} + {\rm H.c.}$ where
${\underline{\xi}_{{\bf q},ab} \equiv (\partial_a u_{{\bf q},b}+\partial_b u_{{\bf q},a})/2}$.

The final term in the Hamiltonian $H_{\rm TLS-TLS}$ characterizes the direct interaction between defects. The static elastic dipole of a defect sources an elastostatic strain field, in analogy with the electrostatic dipole field, that mediates a direct interaction between TLSs given by
\begin{equation}
\label{ }
H_{\rm TLS\mymathhyphen TLS} = \sum_j \sum_{i\neq j} \frac{1}{2} J_{ij} \sigma_{z,i} \sigma_{z,j}.
\end{equation}
The defect-defect coupling strength $J_{ij}$ is determined by separation, relative orientation, and system geometry (to be discussed in detail below). A subleading `flip-flop' contribution to $H_{\rm TLS\mymathhyphen TLS}$ coupling $\sigma_{x,i}$ and $\sigma_{x,j}$, and the effects of retardation have been neglected \cite{Black77}, for further details see Sec. IV C.  

With the theory outlined above, describing the interaction of defects with confined EM and acoustic excitations (see Fig. \ref{HamiltonianPic} b), TLS-induced noise and dissipation in mesoscale systems can be characterized. Such processes are often determined by a large ensemble of TLSs, in such cases the collective influence of the defect bath can be determined statistically. A given system contains a collection of defects, each having a unique energy, coupling, orientation, and position. Statistically, these properties are described by a defect density of states (DDOS) given by $F(\Delta,\Delta_0) = P_D/\Delta_0$ \cite{Phillips87}. When a large number of defects contribute to an observed process the sum $\sum_{j}(...)$ is well-approximated by an ensemble average over defect properties, i.e. $\sum_{j}(...) = V_D \left< \int d\Delta d \Delta_0  F(\Delta,\Delta_0)  ... \right>_{V}$ where angular brackets indicates an average over all possible TLS positions and orientations, and $V_D$ is the $D$-dimensional volume of the system. Generally, this DDOS is position, orientation, and energy dependent, i.e. $P_D \equiv P_D({\bf r}, \hat{n}, E)$, and has units of inverse energy inverse $D$-dimensional volume \cite{Anderson72,Jackle72,Phillips87}. A weak energy dependence of $P_D \propto E^\mu$ has long been suspected as the explanation of the anomalous temperature dependence of the heat capacity in glass, scaling as $T^{1+\mu}$ ($\mu \approx 0.3$), and which recent measurements, directly measuring the distribution of defects in energy, have confirmed \cite{Skacel15}. However, a constant value of $P_D$ ($P_3 \approx 5.45 \times 10^{44} {\rm J/m}^3$ in silica \cite{Golding73,Golding76,Golding76b,Graebner86}) is sufficient to qualitatively and quantitatively explain many phenomena. It should be kept in mind that the distribution of TLS energies, positions and orientations may not be uniform in mesoscopic and/or anisotropic systems. Alternative theories to the TSM attribute the approximately uniform energy dependence of the DDOS to the nature of defect-defect interactions in bulk systems \cite{Yu88,Yu89,Fu89,Leggett91,Vural11,Leggett13}, the behavior of the DDOS is unknown in reduced dimensional systems where defect-defect interactions are modified \cite{Fu89}. It's unknown to what extent crystalline materials may exhibit anisotropy, but systems constructed from materials such as silicon will inevitably have native oxide layers \cite{Al-Bayati90} that will concentrate tunneling states at surfaces.  

\section{defect-phonon and defect-photon interactions in mesoscale systems}

The size and geometry of emerging quantum systems impact the nature of interactions between defects and the EM and acoustic fields, leading to striking transformations of defect dynamics. The interplay of confinement, coherence, and temperature lead to large changes in defect decay-  ($T_1^{-1}$) and dephasing-rates ($T_2^{-1}$), and consequently, noise and dissipation induced by defects is drastically altered as well. In the following sections we discuss each of these aforementioned processes, and how they are modified in mesoscale systems.

\subsection{Defect decay in confined systems}
Defect decay is strongly impacted in mesoscale systems where resonant interactions, contained in $H_{\rm int}$, allow an excited TLS to emit into a number of dispersive, slow-group velocity, or resonator modes. In confined structures the number and nature of these defect decay channels sensitively depends on geometry. This phenomenon is characterized by an excited state decay rate $T_1^{-1}$ given by 
\begin{equation}
\label{T1}
T_1^{-1}(E) =  \sum_{\bf q} \frac{\pi \Delta_0^2}{\Omega_{\bf q} E^2} |\gamma: \underline{\xi}_{\bf q}({\bf r})|^2 \coth \left(\frac{\hbar \Omega_{\bf q}}{2 k_B T}\right) \delta(E-\hbar \Omega_{\bf q})
\end{equation}
that is computed with Eq. \eqref{Hamiltonian} and Fermi's golden rule (see Appendix \ref{T1-Appendix}), where $T$ is the temperature of the phonon field. Equivalently, Eq. \eqref{T1} can be expressed in terms of the phonon density of states (DOS) $g(\Omega)$ by noting that $\sum_{\bf q} \equiv \int_0^\infty d\Omega \ g(\Omega)$. 

In contrast to the result derived from the standard TSM \cite{Phillips87}, the excited state lifetime given in Eq. \eqref{T1} depends upon the position and orientation of the defect, and the mode structure and dispersion of the acoustic field. Such differences are critical to correctly compute $T_1$ in mesoscale systems. However, when the density is spatially uniform a simplification of the decay rate is obtained by averaging Eq. \eqref{T1} over TLS positions and orientations using the identity
\begin{align}
\label{Ident-1}
    {\langle |\gamma: \underline{\xi}_{\bf q}({\bf r}_j)|^2 \rangle_V = 
\frac{1}{V_D} \frac{\Omega_{\bf q}^2}{\rho_D} \sum_\eta \frac{\gamma^2_\eta}{v_\eta^2} e_{{\bf q}\eta}},
\end{align}
where $\rho_D$ is the $D$-dimensional material density (e.g. mass per unit area for $D=2$), and ${e_{{\bf q}\eta}}$ is the fraction of energy of the ${\bf q}$th acoustic mode in the $\eta$-component of the acoustic field (see Appendix \ref{Identity-Appendix}). For example, for plane waves in infinite $D$-dimensional systems, the fractions $e_{{\bf q}\eta}$ equal one when the ${\bf q}$th mode is $\eta$-polarized and zero otherwise, but in compact systems arbitrary modes involve the hybridization of compressional and shear motions, and generally have highly dispersive properties. 

Using Eq. \eqref{Ident-1} the spatial and orientation averaged value of $T^{-1}_1$ is given by
\begin{align}
\label{T1-ave}
\langle T_1^{-1}(E) \rangle_V = & \frac{1}{V_D} \sum_{{\bf q},\eta}  \frac{\pi \Delta_0^2 }{\hbar \rho_D E}  \frac{\gamma_\eta^2}{v_\eta^2}  e_{{\bf q}\eta}  \coth \left(\frac{E}{2 k_B T}\right) \nonumber \\
& \quad \times \delta(E-\hbar \Omega_{\bf q}) \\
\approx  & \frac{1}{V_D} \sum_{\bf q}  \frac{\pi \Delta_0^2 }{\hbar \rho_D E}  \frac{\gamma_\ell^2}{v_\ell^2}   \coth \left(\frac{E}{2 k_B T}\right) \nonumber \\
& \quad \times \delta(E-\hbar \Omega_{\bf q}) \nonumber
\\
\approx & \frac{1}{V_D} g(E/\hbar) \frac{\pi \Delta_0^2 }{\hbar^2 \rho_D E}  \frac{\gamma_\ell^2}{v_\ell^2}   \coth \left(\frac{E}{2 k_B T}\right). \nonumber
\end{align}
The second line follows for materials that satisfy ${(\gamma_\ell/v_\ell)^2 \approx (\gamma_t/v_t)^2}$ (e.g. silica \cite{Anghel07,Footnote-1}) where the identity $e_{{\bf q}\ell}+e_{{\bf q}t} =1$ is used to evaluate the sum on $\eta$, and the third line, expressed in terms of the phonon DOS $g(\Omega)$, follows from the second. When applicable, this approximation greatly simplifies the calculation of the decay rate in complex structures because the acoustic eigenfunctions ${\bf u}_{\bf q}({\bf x})$ are not needed; only the dispersion properties of each acoustic mode are required.   

\subsection{Defect decay in mesoscopic systems}
\begin{figure}[h]
\begin{center}
\includegraphics[width=0.45\textwidth]{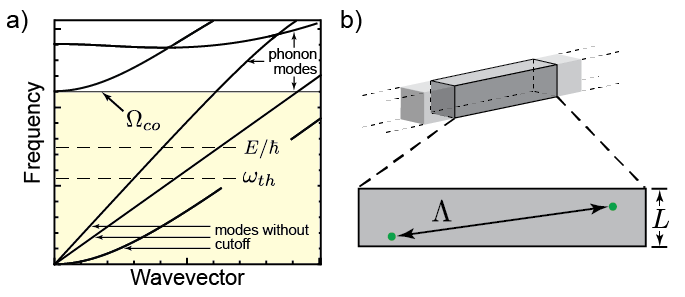}
\caption{Conditions for dimensional reduction: 
a) relevant phonon frequencies less than cutoff $\Omega_{co}$ (yellow region), e.g. $E/\hbar$, frequency of emitted phonons in defect decay, and/or $\omega_{th}=k_B T/\hbar$ thermal frequency, b) mean separation between thermally active defects $\Lambda$ greater than one or more system dimension.}
\label{dimensionalRedux}
\end{center}
\end{figure}

Confined structures support a hierarchy of modes, all but a few of which are cutoff above a system-specific frequency $\Omega_{co}$ (e.g. Fig. \ref{dimensionalRedux}). Defects can emit into all phonon modes at energy E, matching its gap (see Fig. 2). The decay of a defect is dimensionally reduced when the $E < \hbar\Omega_{co}$; in other words, when the emitted phonon wavelength, $\lambda_\eta$, is much smaller than one or more system dimensions. In this limit, illustrated schematically in Fig 5a, the direction of spontaneously emitted phonons is reduced to one of the system's symmetry directions.  This form of dimensional reduction can occur for long wavelength phonons in membranes, microwires, or microtoroids. Despite the reduction in the number of decay channels, the decay rate can be enhanced in mesoscale systems by dispersion and confinement.  
\subsubsection{Geometric enhancement of defect decay in idealized bulk systems}
To draw out the qualitative changes to defect decay as the system dimension is reduced we first consider idealized bulk $D$-dimensional systems ($D\geqslant 1$). 
We define these systems as the lower-dimensional equivalent an infinite $3D$ bulk; they support non-dispersive plane waves that propagate along the system's symmetry directions.
For such systems at low temperatures ($k_B T \ll \hbar \omega_D$ where $\omega_D$ is the Debye frequency) the sum over phonon modes is given by ${\sum_{\bf q} = \int_0^{\omega_D} d\Omega \sum_\eta \frac{V_D}{(2\pi)^D} S_{D-1} \frac{\Omega^{D-1}}{v_\eta^{D}}}$, where $S_{D}$ is the $D$-dimensional unit-hypersphere surface area,  
$\rho_D$ is the $D$-dimensional density, and the acoustic modes are $\eta$-polarized plane-waves with respective sound velocity $v_\eta$.  After averaging over TLS positions and orientations we find the decay rate
\begin{align}
\label{T1,bulk}
\left<T_{1}^{-1}(E)\right>_V  = &  \sum_\eta \frac{\gamma_\eta^2}{v_\eta^{D+2}} \frac{\pi S_{D-1}}{(2\pi)^{D}} \frac{E^{D-2} \Delta_0^2}{\hbar^{D+1} \rho_D} \coth \left( \frac{E}{2 k_B T} \right) \\
 \equiv & \sum_\eta \Gamma^{(D)}_{1,\eta}. \nonumber 
\end{align}
We relate the density in a $D$-dimensional system $\rho_D$ to the bulk density by the formula $\rho_D = \rho L^{3-D}$ where $L$ characterizes the size of the compact dimension(s) (e.g. $L^2$ is the cross-sectional area of a 1$D$ waveguide), and from hereon we drop the explicit label $D$ for $D=3$. Thus, the relative magnitude of $T_1^{-1}$ for infinite $D$-dimensional systems ($D\geqslant1$) is captured by $\Gamma^{(D)}_{1,\eta} \propto S_{D-1}(\lambda_\eta/L)^{3-D}$, showing that the decay rate is geometrically enhanced as the system dimension is reduced since $\lambda_\eta \gg L$. 

The results above give the qualitative behavior for the defect decay in mesoscale systems when the phonon frequency is much less than $\Omega_{co}$. However, in the next section we show that the behavior of $T_1$ varies dramatically from Eq. \eqref{T1,bulk} in compact systems that support dispersive flexural modes, or when the relevant phonon frequencies exceed $\Omega_{co}$.  

\subsubsection{Enhancement of defect decay in waveguides due to dispersion and van-Hove singularities}

\begin{figure}[h]
\begin{center}
\includegraphics[width=0.5\textwidth]{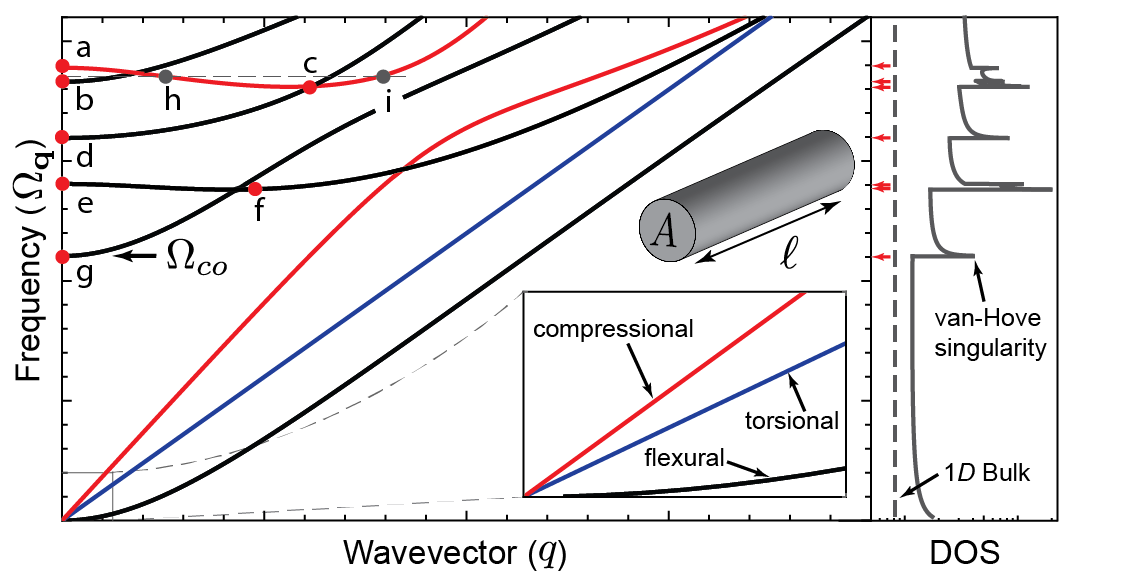}
\caption{Left: Dispersion relations for compressional (red), torsional (blue) and flexural (black) phonon modes in a cylinder. Excitations with zero group velocity are indicated by red points a-g. The branches with points a \& c and e \& f are two examples of modes that have wavevector regions with negative group velocity. The dashed gray line indicates a phonon energy supporting two defect decay channels at the gray points h \& i for a single mode. Inset: System geometry and four fundamental modes without cutoff (two degenerate flexural modes). Right: Phonon density of states in a cylinder (gray) and idealized 1$D$ system (gray dashed). Red arrows indicate frequencies supporting zero group velocity excitations.}
\label{dispRelation}
\end{center}
\end{figure}

Defect decay is enhanced by emission into dispersive and slow group velocity phonon modes. To see this consider an arbitrary translationally invariant system of finite cross section. The phonon mode index ${\bf q}$ in such a system is given by $\{{\bf m},q\}$ where ${\bf m}=(m_1,m_2)$ are indices labeling the eigenfunctions describing the elastic field along the cross section of the waveguide, and $q$ represents the wavevector of the phonon for propagation along the waveguide. Hence, the sum in Eq. \eqref{T1} can be written $\sum_{\bf q} = \sum_{\bf m}\frac{\ell}{2\pi} \int dq$ where $\ell$ is the system length (see Fig. \ref{dispRelation}), and where it should be understood that the mode sum should be cut off so that the number of acoustic modes is finite. Equation \ref{T1} is then evaluated by noting that $\Omega_{\bf q}$ is $\Omega_{\bf m}(q)$ (see Fig. \ref{dispRelation} for the case of a cylinder), the eigenfrequency of the ${\bf m}$th mode evaluated at $q$, and using the delta function identity $\delta(E-\hbar \Omega_{\bf m}(q)) = \sum_j \delta(q-q_{{\bf m}j})|\hbar (\partial \Omega_{\bf m}(q_{{\bf m}j})/\partial q)|^{-1}$. The wavevectors $q_{{\bf m}j}$ are given by all solutions to $E = \hbar \Omega_{\bf m}(q_{{\bf m}j})$; for most modes there is one solution $q_{{\bf m}j}$, but backwards propagating modes (negative group velocity) occur frequently in guided acoustic wave systems and it's possible for such modes to contribute to the decay rate at two values of the wavevector (for example, see the gray points h \& i in Fig. \ref{dispRelation}). Using this identity the integral over $q$ in the mode sum can be directly performed to give 
\begin{align}
\label{T1-guided wave-1D}
\langle T_1^{-1}(E) \rangle_V =  & \frac{1}{A} \sum_{{\bf m},j,\eta}  \frac{\Delta_0^2}{\hbar^2 \rho |v^{{\bf m}j}_{g}|E}  \frac{\gamma^2_\eta}{v_\eta^2} e_{{\bf m}j,\eta}\coth \left(\frac{E}{2 k_B T}\right) \nonumber \\
\approx  & \frac{1}{A} \sum_{{\bf m},j}  \frac{\Delta_0^2}{\hbar^2 \rho |v^{{\bf m}j}_{g}|E}  \frac{\gamma^2_\ell}{v_\ell^2} \coth \left(\frac{E}{2 k_B T}\right)
\end{align}
where $A$ is the waveguide's cross-sectional area, the label $\{ {\bf m}j \}$ is short for $\{ {\bf m},q_{{\bf m}j} \}$, and $v^{{\bf m}j}_{g}$ is the group velocity of the ${\bf m}$th mode evaluated at $q_{{\bf m}j}$. The second line applies to systems where $(\gamma_\ell/v_\ell)^2 \approx (\gamma_t/v_t)^2$, in this approximation the phonon DOS is given by 
\begin{align}
\label{1D-DOS}
  g(\Omega) \!=\! 
  \sum_{\bf m} \!\frac{\ell}{\pi}\! \bigg( \frac{\theta(\Omega\!-\!\Omega_{\bf m}^{min})}{|v^{{\bf m}+}_{g}(\Omega)|} \!+\! \frac{\theta(\Omega_{\bf m}^{co}\!-\!\Omega)\theta(\Omega\!-\!\Omega_{\bf m}^{min})}{|v^{{\bf m}-}_{g}(\Omega)|}
\bigg)
\end{align}
where $\Omega_{\bf m}^{co}$ is the cutoff for the ${\bf m}$th phonon mode (e.g. points a, b, d, e \& g of Fig. \ref{dispRelation}), $\Omega_{\bf m}^{min}$ is the minimum frequency of the ${\bf m}$th dispersion curve (e.g. points b, c, d, f \& g of Fig. \ref{dispRelation}), and $v_g^{\pm}$ is the group velocity in the region where it is positive (+) and where it is negative (-). Notice the divergences (van-Hove singularities) in the DOS for frequencies where the group-velocity vanishes (e.g. points a, b, c, d, e, f \& g in Fig. \ref{dispRelation}).

A similar computation for the defect decay rate in a 2$D$ waveguide can be done. The dispersive properties of acoustic modes in a planar structure is qualitatively similar to a cylinder's shown in Fig. \ref{dispRelation}, with the exception that ${\bf q} = \{m,{\bf k}\}$ where $m$ is a single index labeling the eigenfunctions describing the acoustic field normal to the plane, and ${\bf k}$ is the phonon wavevector in the plane. Using the delta function identity above with $E = \hbar \Omega_m({\bf k}_{mj})$ the decay rate is given by
\begin{align}
\label{T1-guided wave-2D}
\langle T_1^{-1}(E) \rangle_V = & \frac{1}{L} \sum_{m,j,\eta} \frac{\Delta_0^2}{2 \hbar^2 \rho v^{mj}_{p} |v^{mj}_{g}|}  \frac{\gamma^2_\eta}{v_\eta^2} e_{mj,\eta} \coth \left(\frac{E}{2 k_B T}\right) \nonumber \\
\approx & \frac{1}{L} \sum_{m,j} \frac{\Delta_0^2}{2 \hbar^2 \rho v^{mj}_{p} |v^{mj}_{g}|}  \frac{\gamma^2_\ell}{v_\ell^2} \coth \left(\frac{E}{2 k_B T}\right),
\end{align}
where $L$ is the plane thickness, and $v^{mj}_{p}$ and $v^{mj}_{g}$ are the respective phase and group velocities of the $m$th mode evaluated at $|{\bf k}_{mj}|$ (see Appendix \ref{T1-Appendix}). The second line above, where $(\gamma_\ell/v_\ell)^2 \approx (\gamma_t/v_t)^2$, gives 
\begin{align}
\label{2D-DOS}
g(\Omega) \!=\! \sum_{m}\!\frac{A \Omega}{v^m_p(\Omega)}\!\bigg( \frac{\theta(\Omega\!-\!\Omega_m^{min})}{|v^{m+}_{g}(\Omega)|} \!+\! \frac{\theta(\Omega^{co}_m\!-\!\Omega)\theta(\Omega\!-\!\Omega_m^{min}) }{|v^{m-}_{g}(\Omega)|}\bigg),
\end{align}
the phonon DOS in planar waveguides, where the convention of Eq. \eqref{1D-DOS} is used. 

All but a few modes in guided-wave systems are cut off at some finite frequency $\Omega_{co}$ (e.g. see Fig. \ref{dispRelation}), and for defect energies $E \ll \hbar \Omega_{co}$ (or equivalently $\lambda \gg L,\sqrt{A}$), $T_1^{-1}$ is partially described by the results of the previous section. For systems that support flexural modes, however, a more thorough treatment is necessary and Eq. \eqref{T1-guided wave-1D} and \ref{T1-guided wave-2D} must be used. To see how $T^{-1}_{1}$ is modified in reduced dimensional systems that support flexural modes we compute the defect decay rate for cylindrical ($T^{-1}_{1,{\rm cyl.}}$) and planar ($T^{-1}_{1,{\rm pla.}})$ waveguides for $E \ll \hbar \Omega_{co}$ and with stress-free boundary conditions on the acoustic modes. This calculation gives 
\begin{equation}
\label{T1-cyl}
\left.
\begin{array}{r}
  \langle T^{-1}_{1,{\rm cyl.}}  \rangle_V   \\
  \langle  T^{-1}_{1,{\rm pla.}}  \rangle_V   
\end{array} \!\!\!
\right\}
\! \approx \! \frac{\Omega_{\bf q} \gamma_\ell^2}{\hbar \rho v_\ell^2} \coth \frac{\hbar \Omega_{\bf q}}{2 k_B T}
\left\{  \!\!\!
\begin{array}{l}
    \frac{1}{A}\left(\frac{1}{v_B} \! + \! \frac{1}{v_t} \! + \!\sqrt{\frac{2}{v_B \Omega_{\bf q} R}} \right)      \\
      \frac{\Omega_{\bf q}}{L}\left(\frac{1}{v_{pl}^2} \! + \!\frac{1}{v_t^2} \! + \! \frac{\sqrt{3}}{\Omega_{\bf q} v_{pl} L} \right)   
\end{array}
\right.
\end{equation}
where $v_B \equiv \sqrt{Y/\rho}$ and $v_{pl} \equiv v_B (1-\nu^2)^{-1/2}$ are the bar and plate velocity \cite{Rose99} ($Y$ is Young's modulus and $\nu$ is the Poisson ratio), and $R$ is the cylinder radius. The first two terms on the right hand side above represent the contribution to $T^{-1}_1$ from the fundamental compressional (symmetric Lamb wave) and fundamental torsional (shear-horizontal) modes of the cylinder (plate) (Fig. \ref{dispRelation}). Well below cutoff these modes are nondispersive and their dependence on frequency, temperature, and geometry is captured by Eq. \eqref{T1,bulk} (see Inset Fig. \ref{dispRelation}). The last term(s) on the right hand side of Eq. \eqref{T1-cyl} arises from defect decay into the fundamental flexural mode (anti-symmetric Lamb wave) for the cylinder (plate) (see Inset Fig. \ref{dispRelation}). These dispersive modes depend on the system geometry and dominate defect decay into other channels. 

\begin{figure}[h]
\begin{center}
\includegraphics[width=0.45\textwidth]{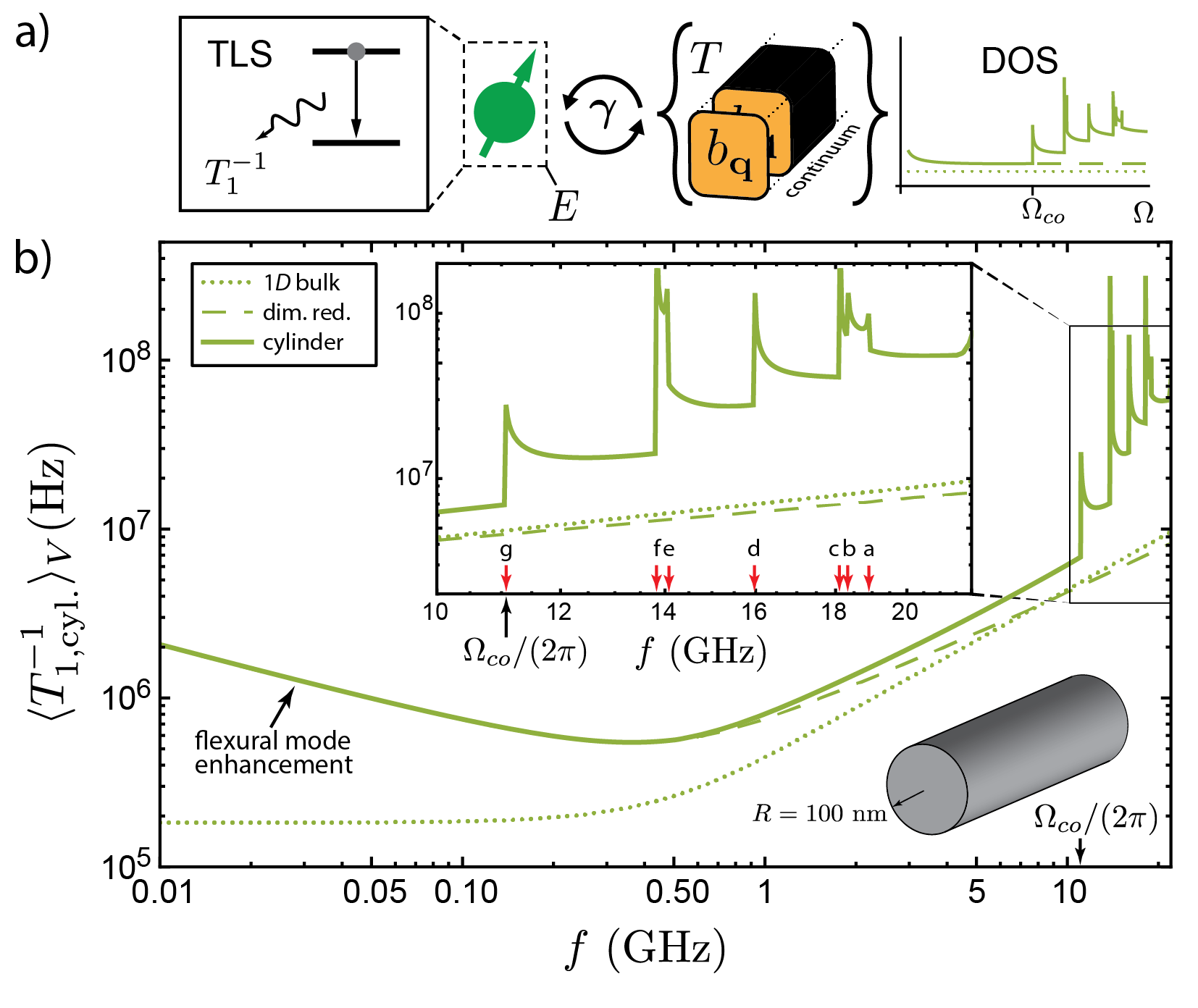}
\caption{a) Illustration of coupling/dynamics leading to defect decay in a waveguide.
b) Defect decay in a silica cylinder ($R=100 \ {\rm nm}$) as a function of defect frequency $f = E/(2\pi \hbar)$ computed using the idealized 1$D$ model (green dotted) Eq. \eqref{T1,bulk}, Eq. \eqref{T1-cyl} for a dimensionally reduced cylinder supporting flexural modes (green dashed), and Eq. \eqref{T1-guided wave-1D} including higher order modes. Red arrows denote frequencies where modes with zero group velocity are supported (see red points of Fig. \ref{dispRelation}). The following parameters are used: $\rho = 2202$ kg/m$^3$, $v_\ell$ = 5944 m/s, $v_t$ = 3764 m/s (and throughout the remainder of the paper), $T=$10 mK, and 
$(\gamma_\ell/v_\ell)^2 \approx (\gamma_t/v_t)^2$ is assumed.}
\label{T1-cyl-fig}
\end{center}
\end{figure}

Also in contrast with bulk systems, a large change in the decay rate occurs for quasi-1$D$ and quasi-2$D$ systems when the energy $E \gtrsim \hbar \Omega_{co}$. For such energies the decay rate can be dramatically enhanced by emission into higher order phonon modes with small group velocity. Equivalently, this enhancement can be interpreted as the result of van-Hove singularities in the phonon DOS (Fig. \ref{dispRelation} \& Fig. \ref{T1-cyl-fig}). Since defect decay into a given channel scales with $v_g^{-1}$, emission into slow group velocity modes can be very large and dominate over other decay processes. Such slow group velocity modes are essentially standing waves transverse to the symmetry direction, and behave similarly to resonator modes. Hence, this enhancement is similar to the Purcell effect \cite{Purcell46,Kleppner81}. 

The effect of emission into flexural and slow group velocity modes on the defect decay rate is illustrated in Fig. \ref{T1-cyl-fig} by comparing $T_1^{-1}$ for an idealized 1$D$ system, a dimensionally reduced cylinder supporting flexural modes (described by Eq. \eqref{T1-cyl}), and a cylinder including all higher order modes. Figure \ref{T1-cyl-fig} shows that $T_1^{-1}$ is dominated by emission into flexural modes at low frequencies, and once $E > \hbar \Omega_{co}$ a large jump in the decay rate is observed at each frequency where slow group velocity excitations are supported.   

\subsubsection{Acoustic Purcell enhanced defect decay in resonators}
Defect decay is strongly modified in high-quality acoustic resonators. In such systems the spectrum of acoustic modes becomes discrete, radically modifying the phonon DOS and leading to Purcell enhancement of the defect spontaneous emission rate \cite{Purcell46,Kleppner81}. 

To understand the acoustic cavity system a golden rule derivation of the decay rate can be carried out that accounts for the coupling of a defect to a discrete set of lossy phonon modes (see Appendix \ref{T1-Appendix}). 
However, if defect-induced losses are the only source of acoustic dissipation in the system then the validity of this treatment requires that the number of defects interacting with a given mode is much greater than 1, i.e. $P \hbar T^{-1}_2 V \gg 1$. In the limit where $P \hbar T_2^{-1} V \lesssim 1$ the composite system will undergo Rabi oscillation (discussed in Appendix \ref{Rabi-Appendix}).

The TLS position- and orientation-averaged decay rate for a defect in a cavity is given by  
\begin{align}
\label{T1,cav}
  \left<T_{1,{\rm cav}}^{-1}(\hbar \omega) \right>_V =  \frac{1}{V_D}\sum_{{\bf q}\eta}& \frac{2\Omega_{\bf q}^2 \Delta_0^2}{\hbar^3 \rho_D \omega }  
   \frac{\gamma^2_\eta}{v_\eta^2} e_{{\bf q}\eta} \coth \left(\frac{\hbar \omega}{2 k_B T}\right)
  \nonumber \\ &  \times  \frac{\Gamma_{\bf q}}{(\omega^2 - \Omega_{\bf q}^2)^2 + \omega^2 \Gamma_{\bf q}^2}
\end{align}
where $\Gamma_{\bf q}$ is the decay rate of the ${\bf q}$th phonon mode. 
A few remarks are necessary regarding Eq. \eqref{T1,cav}. Without the adoption of an explicit cutoff, either given by the defect size or the Debye frequency, the sum in the defect decay rate above diverges. Consequently, Eq. \eqref{T1,cav} has a potentially large cutoff-dependent contribution. Such a cutoff dependence occurs in the theoretical treatment of spontaneous emission of atoms embedded in absorbing dielectrics where it is attributed to non-radiative decay through the near field \cite{Scheel99}. The cutoff-dependent component of Eq. \eqref{T1,cav} is contained entirely in $ \left<T_{1,{\rm cav}}^{-1}(\hbar \omega = 0) \right>_V$ which suggests that it is the elastic analog of non-radiative decay. It is necessary to consider the nature of phonon decay in a given system to determine whether the cutoff-dependent term should be retained (see Appendix \ref{T1-Appendix}).
\begin{figure}[h]
\begin{center}
\includegraphics[width=0.45\textwidth]{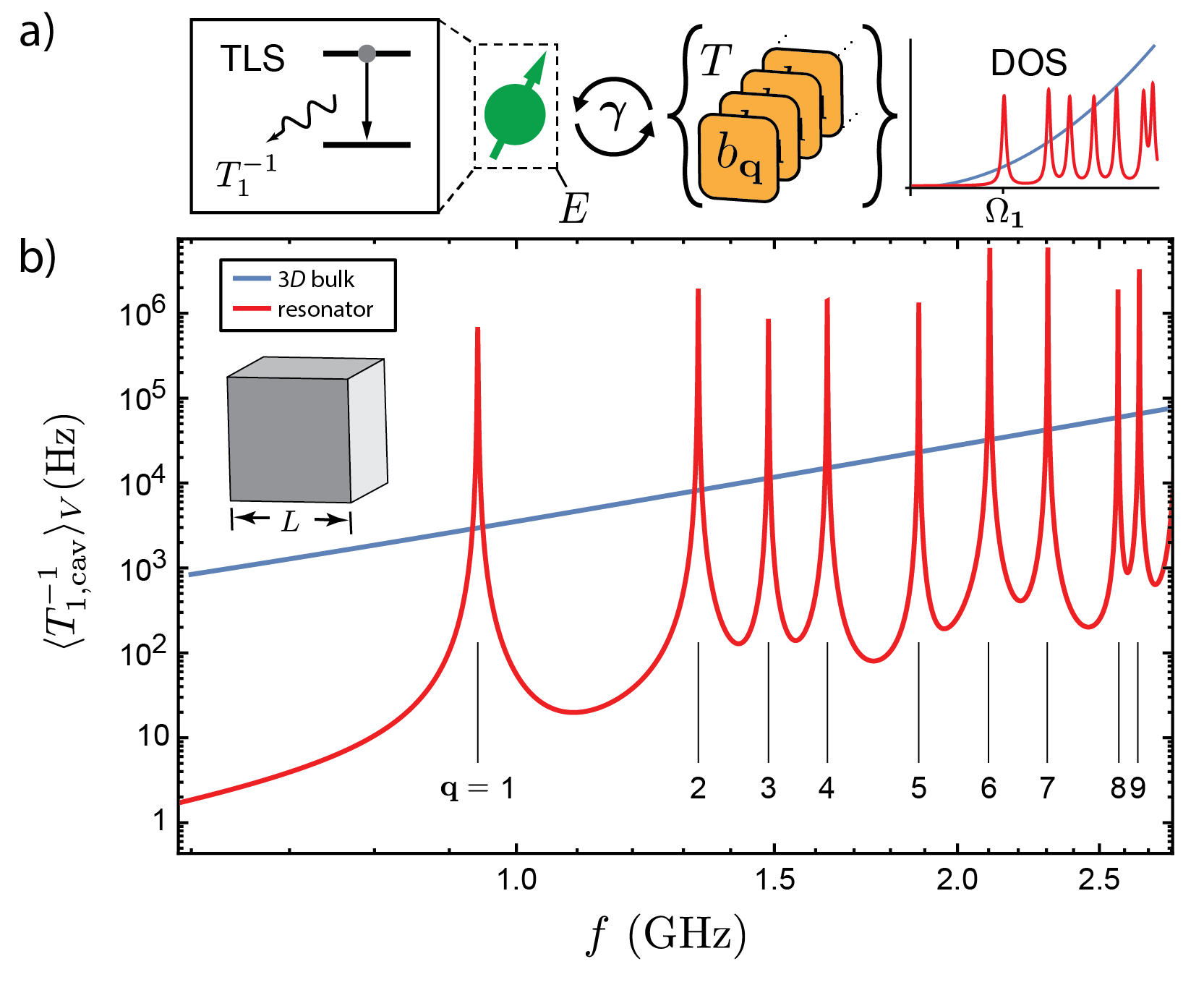}
\caption{a) Illustration of coupling/dynamics leading to defect decay in a resonator.
b) Decay defect decay as a function of $f = E/(2\pi \hbar)$ in a 3D bulk (blue) and a 4 $\mu$m cubic silica resonator (red). The resonator is defined with periodic boundary conditions applied across parallel faces. The zero energy contribution to Eq. \eqref{T1,cav} has been subtracted, and $T=$ 5 mK.}
\label{T1-resonator}
\end{center}
\end{figure}

To interpret this result it is useful to categorize the defects in the cavity system into two classes: {\it resonant} defects have energies within the linewidth of the cavity's acoustic resonances, and all remaining defects are deemed {\it non-resonant}.  Eq. \eqref{T1,cav} predicts highly suppressed decay for non-resonant defects as compared to a bulk medium, whereas the decay rate of resonant defects is enhanced. In a high-finesse acoustic cavity the decay rate for a resonant defect with $E \approx \hbar \Omega_{\bf q}$ is dominated by a single term in the sum over modes in Eq. \eqref{T1,cav} (assuming nondegeneracy) where the Lorentzian factor reduces to $\sim 2/\pi\Gamma_{\bf q}$. A relationship between the decay rate for defects in a bulk system and resonant defects in a cavity can be derived by expressing Eq. \eqref{T1,cav} in terms of the phonon wavelength and acoustic quality factor $Q_{\bf q} \equiv \Omega_{\bf q}/\Gamma_{\bf q}$ giving 
\begin{equation}
\label{Purcell}
\left< T_{1,{\rm cav}}^{-1}(\hbar \Omega_{\bf q}) \right>_V  \approx \sum_\eta
\frac{1}{2\pi^2}\frac{\lambda^3_\eta}{V} Q_{\bf q} e_{{\bf q}\eta}  \Gamma_{1,\eta}
\end{equation}
for $D=3$.
The prefactor $(\lambda^3_\eta/ 2\pi^2 V) Q_{\bf q} e_{{\bf q}\eta}$ gives the acoustic analog of Purcell enhancement, and indicates that for low order modes, where $\lambda_\eta^3 \sim V$, that the decay rate for resonant defects is dramatically increased by a factor $Q_{\bf q}$. 

\subsubsection{Defect decay through photon emission}
In many scenarios defect relaxation is dominated by decay through phonon emission. However, in high-quality EM waveguides or resonators defect decay via photon emission may become important. In such cases the defect decay rate can be obtained from the results of the preceding sections by the replacement 
\begin{align}
\label{}
 \langle |\gamma: \underline{\xi}_{\bf q}({\bf r}_j)|^2 \rangle_V  \to  & \Omega_{\bf q}^2 \langle |{\bf d} \cdot {\bf E}_{\bf q}({\bf r}_j)|^2 \rangle_V =  \frac{ \Omega_{\bf q}^2 |{\bf d}|^2}{3 V_D \varepsilon_0 \varepsilon_{\bf q}^{\rm eff}}.
\end{align} 
Above, ${\bf E}_{\bf q}$ is an orthonormal eigenfunction of Maxwell's equations satisfying $\nabla \times \nabla \times {\bf E}_{\bf q}({\bf x})  = \varepsilon({\bf x})  (\Omega_{\bf q}/c)^2 {\bf E}_{\bf q}({\bf x})$ and $\int_{V^{\rm em}} d^3 x \  \varepsilon_0 \varepsilon({\bf x}) {\bf E}^*_{\bf q}({\bf x}) {\bf E}_{\bf q'}({\bf x}) = \delta_{\bf q q'}$, $\varepsilon_0$ and $\varepsilon({\bf x})$ are the free space and relative permittivity, and $c$ is the speed of light in vacuum. 
$V^{\rm em}$ is the EM mode volume which is generally different from the volume containing the defects $V$, e.g. this occurs in hollow EM resonators or waveguides that contain oxide surface layers or small dielectric samples. $\varepsilon^{\rm eff}_{\bf q}$ is an effective permittivity defined as $(\varepsilon_{\bf q}^{\rm eff})^{-1} \equiv \varepsilon_0 \int_V d^3 x \  |{\bf E}_{\bf q}({\bf x})|^2$ (note the spatial integral is over $V$). For systems where $V \ll V_{\rm em}$ $\varepsilon_{\bf q}^{\rm eff} \sim \varepsilon(V_{\rm em}/V)$ where $\varepsilon$ is the relative permittivity of the region not containing defects. 

\subsection{Defect dephasing in mesoscale systems}
The nature of defect dephasing depends sensitively on the interplay of geometry and defect concentration. Dephasing is in part the result of resonant processes (as described above) but it is often dominated by perturbations of a defect's energy originating from defect-defect interactions mediated by the elastic field \cite{Black77}. In general, these interactions are described by $H_{\rm TLS\mymathhyphen TLS}$, as well as an additional `flip-flop' contribution that permits a direct transfer of energy between two mutually-resonant defects. However, with energies distributed over a wide range, TLSs do not have an innate energy scale, unlike atoms and nitrogen-vacancy centers, and hence, the number of mutually resonant TLSs at a given energy is insignificant compared to the total sum of defects. For this reason, `flip-flop' interactions do not significantly contribute to dephasing, and the interaction Hamiltonian $H_{\rm TLS\mymathhyphen TLS}$ can be used to estimate $T_2$.

Defect dephasing mediated by direct defect interactions can be understood by adding $H_{\rm TLS\mymathhyphen TLS}$ to the Hamiltonian for the non-interacting defects. Some rearrangement shows that the defect energy can be redefined as 
\begin{equation}
\label{energyShift}
E'_j = E_j + \sum_{i\neq j} J_{ij} \sigma_{z,i},
\end{equation}
which is shifted by an amount determined by the configuration of all remaining TLSs in the system. 
$E'_i$ becomes a function of time when the $i$th defect's neighbors undergo dynamical spin flips. 
Hence, the collective oscillation of an ensemble
of defects with energy $E'$, excited by a strong monochromatic
acoustic pulse, dephases in time when the energy of each defect in the ensemble
randomly hops from its initial value as it's neighbors undergo spin flips induced by thermal fluctuations. 

This process of `spectral diffusion' arises from thermally active defects, with $E < k_BT$ and a concentration $n \sim P(k_B T) k_B T$, that can absorb and emit thermal phonons. Defects with $E \gg k_BT$ are frozen in their ground state, and thus contribute a time-independent shift to the defect energy.   
The concentration of thermally active defects defines a spectral diffusion length scale $\Lambda \sim [P(k_B T) k_B T]^{-1/3}$. For example,  $\Lambda \approx 237 \ {\rm nm} (\frac{10 {\rm mK}}{T})^{1/3}$ for silica \cite{Graebner86}. When one or more system dimensions is much smaller than $\Lambda$ spectral diffusion is dimensionally reduced (see Fig. \ref{dimensionalRedux}); a scenario achievable with standard fabrication and cryogenic capabilities. 
This dimensional reduction is accounted for in the coupling parameter $J_{ij}$ whose magnitude is set by the system dimension and the separation between the $i$th and $j$th defects $r_{ij} \equiv |{\bf r}_i-{\bf r}_j|$. 

Direct interactions between defects are mediated by the static strain field (the elastic equivalent of the electrostatic dipole field). In 3$D$ systems the coupling between defects scales as $J_{ij} \propto r_{ij}^{-3}$ (for more details see the Appendix of \cite{Black77}).  In 2$D$ $J_{ij}$ falls off as $r_{ij}^{-2}$, enhancing defect-defect interactions, but surprisingly, the coupling is completely local in idealized 1$D$ systems scaling as $J_{ij} \propto \delta(r_{ij})$. 

To understand the true spatial scaling of $J_{ij}$ in a non-idealized 1$D$ system we derived an approximate expression for the static elastic strain field in a microwire with a finite cross-section. In such a system the strain field is represented as an infinite sum of eigenfunctions (e.g. Bessel functions) that describe the dependence of the elastic field perpendicular to the symmetry direction \cite{Fu89}. The dependence of the strain field along its symmetry direction scales as $e^{- \frac{x}{R} |z -z'|}$ where $x$ is a pure number, e.g. a Bessel function zero, $R$ represents the system `radius', and $|z-z'|$ is the separation along the wire between two defects. 
The parameter $x$ and the separation $|z-z'|$ have minimum values respectively of order 1 and $\Lambda$. Therefore, the static strain field in a non-idealized 1$D$ system is exponentially suppressed in the limit that $\Lambda/R \gg 1$, yielding the coupling $J_{ij} \sim  \frac{x}{2 \rho_{1} R} e^{- \frac{x}{R} r_{ij}}$ \cite{Fu89}, and reduces to idealized result $J_{ij} \propto \delta(r_{ij})$ in the $R\to 0$ limit.

A qualitative understanding of spectral diffusion in reduced dimensional systems is obtained from the dependence of $J_{ij}$ on $r_{ij}$ by following the treatment of Black and Halperin \cite{Black77} and Phillips \cite{Phillips87}. The root-mean square variation of the energy of a given defect due to spin flips of its neighbors $\Delta E$ is estimated by replacing $r_{ij}$ with $\Lambda$ in $E'_i-E_i$ (appearing in $J_{ij}$ of Eq. \eqref{energyShift}). The dephasing time-scale arising from spectral diffusion, the time for defects excited by a monochromatic pulse to begin oscillate out of phase, is approximately given by $T^{'}_2  \sim \hbar/\Delta E$
\begin{equation}
\label{T2-tempScaling}
\frac{1}{T^{'}_2} \sim   
 \left\{
\begin{array}{ll}
     \frac{1}{ \hbar \rho} C_{\rm rms} P(k_B T) k_B T            & D = 3  \\
     \frac{1}{
     \hbar \rho_2} C^{(2)}_{\rm rms} [P(k_B T) k_B T]^{2/3} & D = 2 \\
     \frac{x}{
     \hbar \rho_1 R} C^{(1)}_{\rm rms} \exp\{- \frac{x}{R} [P(k_B T) k_B T]^{-1/3}\}                & D = 1  \\
\end{array} 
\right. 
\end{equation} 
where $C^{(D)}_{\rm rms}$, of order $\gamma^2/v^2$, is defined by the tensor structure of the static strain field \cite{Black77}.
Recalling the definition $\rho_D = \rho L^{3-D}$,
the results above show that the increase in the defect-defect coupling enhances ${T'}^{-1}_2$ by $\Lambda/L$ over the 3$D$ result in planar systems. In contrast, the interaction between defects is effectively local in 1$D$ systems, leading to the exponential suppression of spectral diffusion. However, since defect dephasing is also augmented by resonant phonon processes the total dephasing rate is given by $T^{-1}_2 = \frac{1}{2}T_1^{-1} + T_2^{'-1}$, meaning that $2 T_1$ and $T_2$ are equal in 1D systems. 

Defect dephasing has been measured in phonon echo experiments, finding $T_2 \approx 14-20 \ \mu$s for 0.68 - 1.2 GHz phonons at 20 mK in silica \cite{Golding76,Enss96a,Enss96b}. In the remainder of the paper we use the measured value of $T_2 = 14 \ \mu$s (for all $D$) and extrapolate to other temperatures using Eq. \eqref{T2-tempScaling}. However, this extrapolation overestimates the late time value of $T_2$ since the phonon echo experiments in \cite{Golding76,Enss96a,Enss96b} were performed in the short time limit before the dephasing rate reached its steady-state value \cite{Black77}. 
 
\subsection{Defect-driven noise in mesoscale systems}
In this section we investigate geometry-, dimension, and scale-induced transformations of radio frequency (RF) noise generated by defects. Such noise has been identified as a key limitation in a number of quantum systems \cite{Simmonds04,Astafiev04,Martinis05,Gao07,Wang09,Lindstrom09,Macha10,Pappas11,Sage11,Khalil11,Burnett14} and is generated when a defect's electric dipole moment
${\boldsymbol{\mathcal P}_j (t) \equiv {\bf d}_j\left(\frac{\Delta_{0j}}{E_j} \sigma_{x,j} + \frac{\Delta_j}{E_j} \sigma_{z,j} \right)}$
is stochastically driven by a thermal bath of phonons. 
This physics is phenomenologically described by the Bloch equations with defect relaxation- and dephasing-rates as inputs. The power spectrum of these dipole fluctuations characterizes the electromagnetic noise arising from defects, and can be computed from the Fourier transform of the two-time dipole correlation function given by
\begin{equation}
\label{ }
{\bf S}_{ij}(\omega) = \int_{-\infty}^{\infty} d \tau \ e^{i \omega \tau} \langle  \boldsymbol{\delta \mathcal{P}}_i(t) \boldsymbol{\delta\mathcal{P}}_{j}(t-\tau) \rangle
\end{equation}
where $\boldsymbol{\delta \mathcal{P}}_j \equiv \boldsymbol{\mathcal{P}}_j - \langle \boldsymbol{ \mathcal{P}}_j \rangle$ and $\langle .. \rangle$ denotes expectation value. 
 
We approximate the power spectrum of a single defect's dipole fluctuations with the quantum regression theorem (QRT) \cite{MandelWolf}, a tractable method to obtain a correlation function that is consistent with the Pauli operator algebra and statistical mechanics. Such a computation (see Appendix \ref{Correlation-Appendix}) gives
\begin{align}
\label{PS}
{\bf S}_{ij}(\omega) =  \delta_{ij} {\bf d} {\bf d} 
\bigg[  & \frac{\Delta_{0}^2}{E^2} \bigg( p_e(E) \frac{2 T_2}{1+T_2^2(E/\hbar+\omega)^2} 
\\
& \quad +p_g(E) \frac{2 T_2}{1+T_2^2(E/\hbar-\omega)^2} \bigg) \nonumber \\
& +\frac{\Delta^2}{E^2}\sech^2 (E/2 k_B T) \frac{2 T_1}{1+T_1^2 \omega^2}   \bigg], \nonumber 
\end{align} 
where $p_e(E)$ ($p_g(E)$) is the probability for a defect of energy $E$ to be in the excited (ground) state \cite{Shnirman05,Constantin09}. 

There are two physical mechanisms leading to the noise: resonant and relaxation processes respectively contributing ${\bf S}^{res}_{ij}(\omega)$ and ${\bf S}^{rel}_{ij}(\omega)$ to the power spectrum. ${\bf S}^{res}_{ij}(\omega)$ is made up of the terms proportional to $\Delta_0^2$ in Eq. \eqref{PS} and arises primarily from spin flips of defects associated with the absorption and emission of phonons with energies matching the defect energy. Whereas, ${\bf S}^{rel}_{ij}(\omega)$ is given by the term proportional to $\Delta^2$ in Eq. \eqref{PS} and arises from a stochastic modulation of the defect energy levels by phonons leading the defects to reradiate. 

Although we find a system-independent functional form for the power spectrum, quantitative and qualitative changes in the defect-induced noise result from the dependence of the defect relaxation and dephasing rates on the system size and geometry. To highlight the contrasting behavior of noise in mesoscale systems we consider noise from a single defect in a resonator, and the noise from an ensemble of defects in idealized bulk systems and resonators. 
 
\subsubsection{Acoustic Purcell modulation of noise from a defect in a resonator}
\begin{figure}[h]
\begin{center}
\includegraphics[width=3.3in]{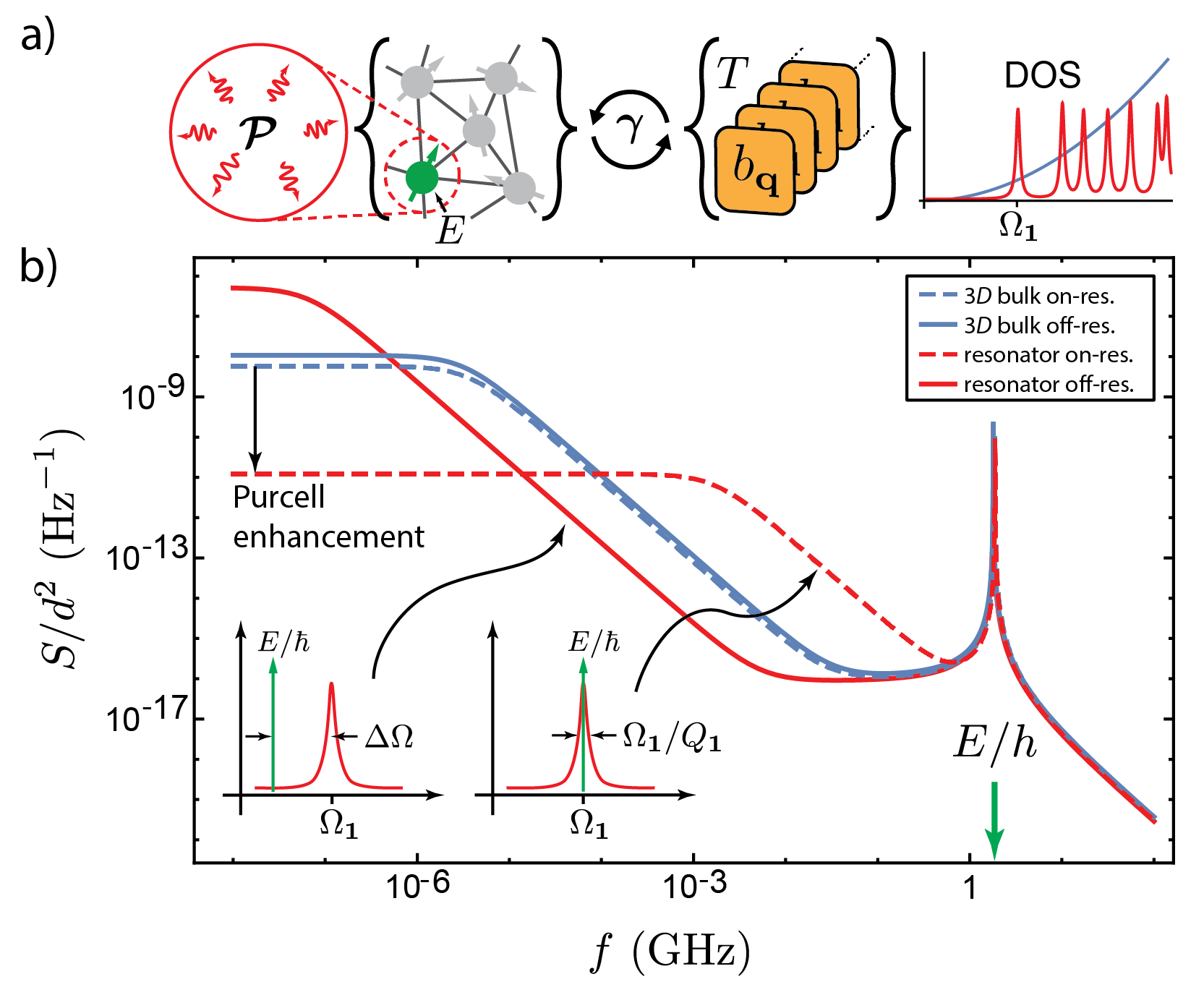}
\caption{a) Illustration of coupling/dynamics of defect-induced RF noise in a resonator. b) Power spectrum of dipole fluctuations of a single defect on-resonance (red-dashed line) and $-0.04 = \Delta \Omega/\Omega_1$ fractionally detuned from (red line) a 2 $\mu$m cubic silica resonator's fundamental acoustic mode at frequency $\Omega_1 = (2\pi) 1.882$ GHz. Periodic boundary conditions are applied to each face, the $Q$ of the fundamental mode is taken to be $1882$, and the temperature is 10 mK. The power spectrum for the same defects in 3$D$ bulk is displayed in blue for comparison (blue-dashed, resonance frequency) and (blue line, detuned frequency}
\label{PS-Single-Defect-Cavity}
\end{center}
\end{figure}

Next, we examine the EM power spectrum of a single defect that interacts strongly with an acoustic resonator.  Note that the active defect is one of coupled ensemble of defects, as diagrammed in Fig \ref{PS-Single-Defect-Cavity}. The noise from a single defect in this resonator is strongly affected by Purcell enhancement. Since $T_2 \leqslant 2 T_1$ Eq. \eqref{PS} shows that the magnitude of the noise at low-frequencies is set by $T_1$, and therefore a sharp contrast in the magnitude of the power spectrum will occur for defects that are on- and off-resonance with an acoustic mode of a high-quality resonator.  
This modulation is illustrated in Fig. \ref{PS-Single-Defect-Cavity} where the dipole power spectrum for two defects in a resonator is compared: one defect is resonant, and the other is fractionally detuned by $- 4 \%$ from the fundamental acoustic mode. Figure \ref{PS-Single-Defect-Cavity} shows that this small fractional detuning ($-82$ MHz) produces shifts in the magnitude of the low-frequency noise of nearly 5 orders of magnitude. This result may point toward new techniques to engineer noise in quantum information systems. Recently, strain tuning of TLS frequencies, of order 100 MHz, has been demonstrated in qubits \cite{Grabovskij12}, suggesting that a large modulation of the RF noise from TLSs could be achieved in quantum information systems formed into high quality acoustic resonators. 

\subsubsection{Geometric modification of the noise from defect ensembles in bulk systems}
Geometric modifications of defect dynamics reshapes the noise from ensembles of defects in reduced dimensional systems. To examine the qualitative features of such reshaping, we consider the power spectrum arising from an ensemble of defects in idealized bulk systems of various dimensions. When a large number of defects contribute to the RF noise the total power spectrum can be computed from the ensemble average of Eq. \eqref{PS} over the defect properties
$
{\bf S}_{tot}(\omega) = \sum_{i} {\bf S}_{ii}(\omega) \approx
V_D \left< \int d\Delta d \Delta_0  F(\Delta,\Delta_0) {\bf S}_{ii}(\omega) \right>_{V}.
$
This approximation is valid in the weak coupling limit when the fluctuations of any two defects is uncorrelated to leading order. 

First, we analyze the noise arising from resonant absorption ${\bf S}^{res}_{tot}(\omega)$. To compute the defect ensemble average, we take $F(\Delta,\Delta_0) = P_D(E)/\Delta_0$ for simplicity, but note that a variety of power spectra are obtained by using more general DDOS \cite{Shnirman05}. After a change of variables to `polar' coordinates, i.e. $\Delta = E \cos \phi$ and $\Delta_0 = E \sin \phi$ recalling $E = \sqrt{\Delta^2 + \Delta_0^2}$, and evaluating the $\phi$-integral, ${\bf S}^{res}_{tot}(\omega)$ is given by 
\begin{align}
\label{Res-Noise}
{\bf S}^{res}_{tot}(\omega) \approx & \frac{2 |{\bf d}|^2}{3}{\bf I} \ V_D  
\int_{0}^{\infty} dE  
\frac{\left< P_D(E) \right>_V p_g(E) T_2}{1+ (\omega - E/\hbar)^2T_2^2} 
\nonumber \\
\approx &  \frac{\pi \hbar |{\bf d}|^2}{3}{\bf I}  \ V_D  \left< P_D(\hbar \omega) \right>_V p_g(\hbar \omega).
\end{align} 
Here ${\bf I}$ is the $3\times3$ identity matrix, a negligible contribution from the anti-resonant term has been dropped and the second line holds for $\omega T_2 \gg 1$. This result shows that the noise arising from resonant processes scales with energy dependence of the DDOS (since $p_g$ varies between 0.5 and 1) and the system size. 

\begin{figure}[h]
\begin{center}
\includegraphics[width=3.3in]{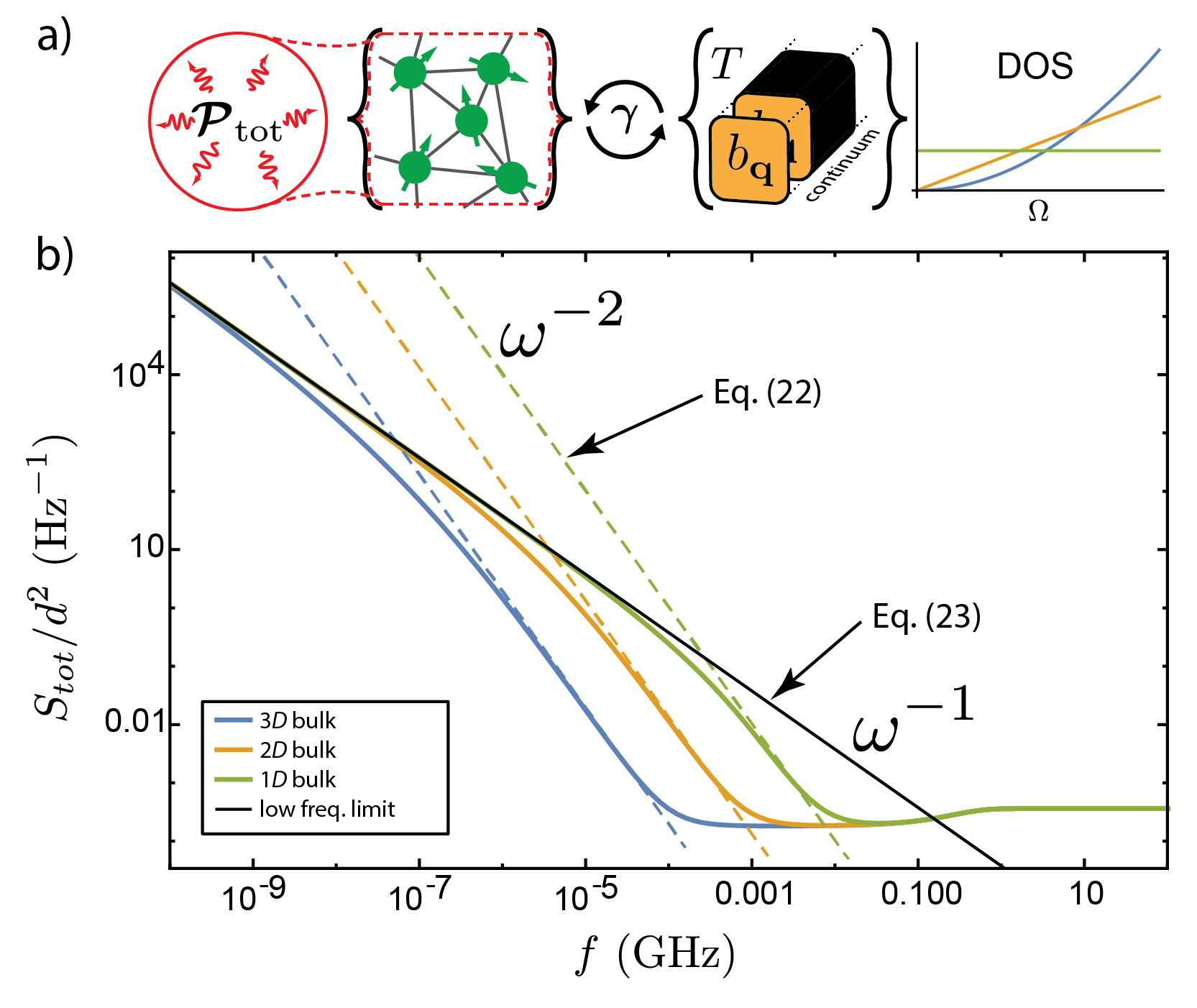}
\caption{a) Illustration of coupled system leading to RF noise. b) Power spectrum for dipole fluctuations from an ensemble of defects in 1, 2 and 3$D$. The compact dimension(s) and the temperature are respectively taken to be $50$ nm and 10 mK ($\omega_{th} = 208$ MHz). The volume of each systems is fixed to $(10 \ \mu{\rm m})^3$ so that each system possesses the same number of defects. We adopt a uniform DDOS, given in Sec II., and adopt the defect properties of silica to compute the ensemble average.}
\label{Ensemble-Noise-Bulk}
\end{center}
\end{figure}

The relaxation component of the total power spectrum is given by 
\begin{align}
\label{Rel-Noise}
{\bf S}^{rel}_{tot}(\omega) \approx  \frac{2 |{\bf d}|^2}{3}{\bf I} \ V_D & \left< 
\int d\Delta d \Delta_0 \frac{P_D(E)}{\Delta_0} 
\frac{\Delta^2}{E^2}\sech^2 \left(\frac{E}{2 k_B T}\right) \right.
\nonumber \\ & \quad
\left.
\times \frac{T_1}{1+T_1^2 \omega^2} 
\right>_V.
\end{align}
Noting that $T_1$ takes a minimum value, $T_{\rm 1, min}$, when $\Delta_0 = E$, a change of variables to $(E,\phi)$, and expressing $T_1$ as $T_{1,{\rm min}} E^2/\Delta_0^2$ allows the $\phi$-integral to be done analytically. The resulting expression is complicated so we present the resulting power spectrum in the high- and low-frequency limits. For high frequencies, i.e. $\omega T_{1,{\rm min}}(k_B T) \gg1$, the idealized bulk system power spectrum due to relaxation processes reduces to  
\begin{align}
\label{rel-Abs-HF-noise}
{\bf S}^{rel}_{tot}(\omega) \approx  \frac{2   |{\bf d}|^2{\bf I} \ V_D P_D(k_B T)}{9 \omega^2} & 
 \sum_\eta \frac{\gamma_\eta^2}{v_\eta^{D+2}} \frac{\pi S_{D-1}}{(2\pi)^{D}} \nonumber \\
 & \times 
\frac{(2 k_B T)^{D+1}}{\hbar^{D+1} \rho_D} \mathcal{I}_{D+\mu}
\end{align} 
where $\mathcal{I}_m \equiv  \int_0^\infty d y \ y^{m}  \sech^2 y \coth y$, and the DDOS is proportional to $E^\mu$ (as discussed in Sec. III). Up to pure numerical factors, the noise from Eq. \eqref{rel-Abs-HF-noise} is enhanced for $D<3$ by a factor $(\lambda_{th}/L)^{3-D}$ where $\lambda_{th} = 2\pi \hbar v_\eta/(k_B T)$ is the thermal wavelength, see Fig. \ref{Ensemble-Noise-Bulk}. 

At low frequencies relaxation absorption results in $1/f$-noise given by
\begin{align}
\label{Rel-Noise-LF}
{\bf S}^{rel}_{tot}(\omega) 
 \approx & \frac{\pi  |{\bf d}|^2{\bf I} \ V_D}{3 \omega} P_D(k_B T) k_B T c_\mu
\end{align} 
where $c_\mu = \int_0^\infty d y \ y^{\mu}  \sech^2 y$. Since the product of $V_D P_D$ is independent of $D$, the low-frequency behavior of the noise is universal (see Fig. \ref{Ensemble-Noise-Bulk}). 

\subsubsection{Thermal suppression of noise from defect ensembles in resonators}
\begin{figure}[h]
\begin{center}
\includegraphics[width=3.3in]{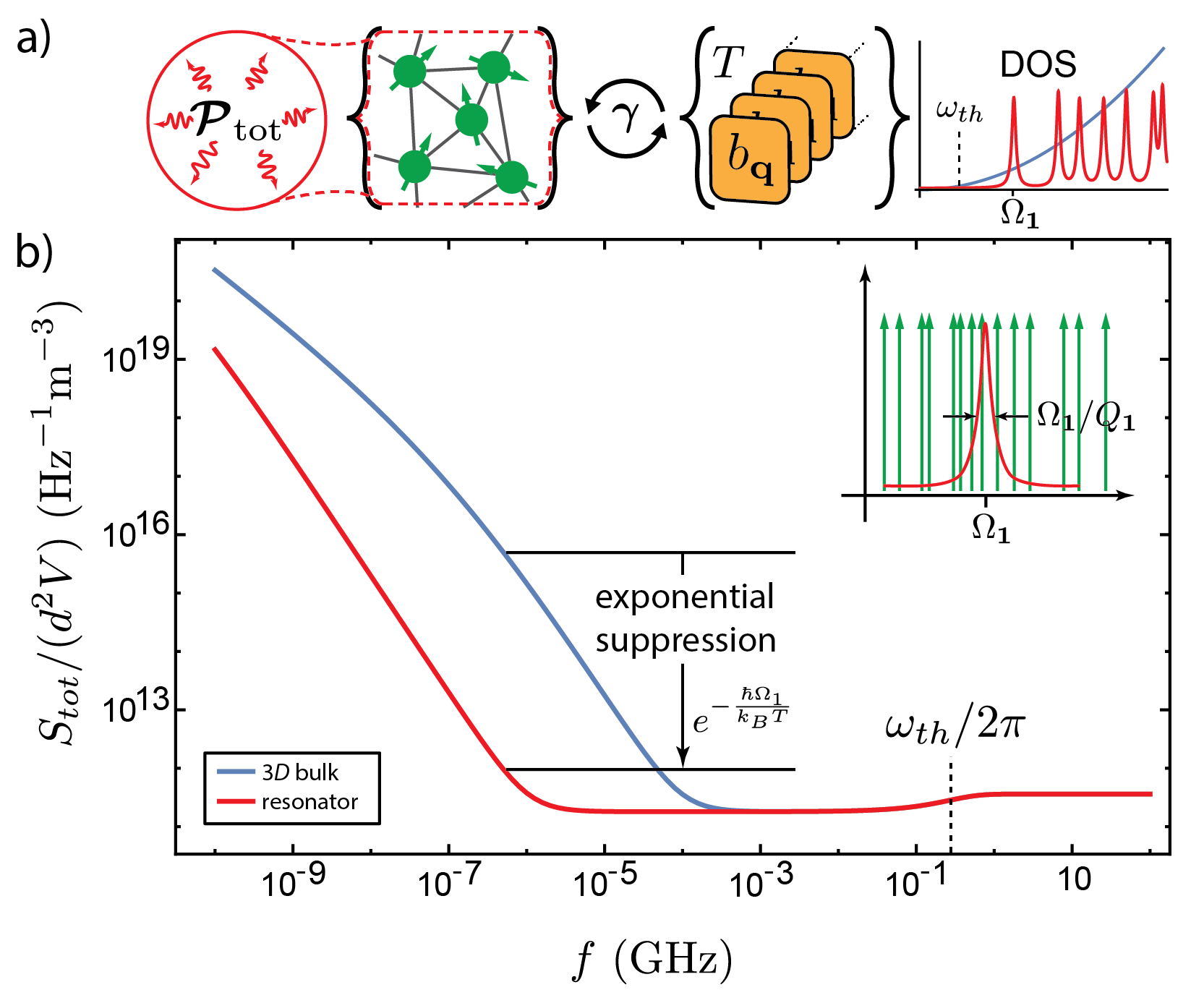}
\caption{Power spectrum per unit volume from a defect ensemble in a 3$D$ bulk (blue) and resonator (red) system at $10$ mK. Material properties of silica are used and periodic boundary conditions are implemented on a cube of side $L = 1\ \mu$m to model the resonator. }
\label{PS-Ensemble-Cavity}
\end{center}
\end{figure}

In this section, we illustrate how the power spectrum from an ensemble of defects in a resonator is exponentially suppressed at low temperatures. In resonators, ${\bf S}^{res}_{tot}(\omega)$ is well-approximated by Eq. \eqref{Res-Noise}, but, in contrast the noise from relaxation absorption depends sensitively on the phonon DOS. For realistic values of the acoustic mode decay rate the high-frequency limit applies over a broad range of frequencies, allowing a Taylor expansion in large $\omega T_1$ to be taken in the integrand of Eq. \eqref{Rel-Noise}. Given $T_1$ for defects in an acoustic resonator the integral in Eq. \eqref{Rel-Noise}  is approximately given by 
\begin{align}
\label{Rel-Noise-HF-Cav}
{\bf S}^{rel}_{tot,cav}(\omega) \approx  \frac{4\pi |{\bf d}|^2{\bf I}\ V_D}{9 \omega^2} & 
\sum_{\bf q} 
 \frac{\left< P_D(\hbar \Omega_{\bf q}) |\gamma: \underline{\xi}_{\bf q}({\bf r})|^2 \right>_V}{\Omega_{\bf q} \sinh \left(\frac{\hbar \Omega_{\bf q}}{k_B T}\right)}. 
\end{align}
Unlike bulk systems, possessing a continuum of phonon modes to thermally drive defect fluctuations, resonators have a gapped, discrete spectrum where Langevin forcing is concentrated near cavity resonances. As a consequence, the thermo-mechanical motion driving defect noise can be frozen out at low temperatures (i.e. $k_B T < \hbar \Omega_1$, where $\Omega_1$ is the frequency of the resonator's fundamental mode) leading to exponential suppression of ${\bf S}^{rel}_{tot,cav}(\omega)$. This suppression is illustrated in Fig. \ref{PS-Ensemble-Cavity} where the power spectral density from an ensemble of defects in a cubic resonator made of silica is compared to a bulk system. The frequency of the fundamental mode, $\Omega_1 = (2\pi)3.7$ GHz, was chosen to be much larger than the thermal frequency $\omega_{th} = 208$ MHz (for $T= 10$ mK) in order to freeze out the resonator's thermo-mechanical motion and exponentially suppress the noise.   

\subsubsection{Scaling of frequency noise from strongly-interacting defects in reduced dimensions}
The previous sections focused on RF noise generated by ensembles of weakly coupled defects. 
However, recent measurements suggest that strongly interacting defects are an important source of frequency noise in superconducting circuits \cite{Burnett14}.  These measurements are explained by a generalized tunneling state theory proposed by Faoro and Ioffe \cite{Faoro15} which predicts a frequency noise power spectrum $S_\nu(\omega)$ proportional to $T_2/\omega$ at low field intensities, and proportional to $\sqrt{T_2/T_1}[\sqrt{\mathcal P} \omega]^{-1}$ at high RF power $\mathcal{P}$ \cite{Faoro15}. Given the dependence of $S_\nu(\omega)$ on $T_1$ and $T_2$ the frequency-noise arising from strongly interacting defects is sensitive to the system geometry. The qualitative behavior of the power spectrum can be derived at low intensities by using the scaling of the power spectrum with $T_2$ for $D\leqslant3$
\begin{equation}
\label{}
S_\nu(\omega) 
\propto \frac{1}{\omega}
 \left\{
\begin{array}{ll}
      \rho T^{-1-\mu}  & D = 3  \\
     \rho_2 T^{-2(1+\mu)/3} & D = 2 \\
     T_1                      & D = 1  \\
\end{array} 
\right. 
\end{equation}
where $T_2 = 2 T_1$ has been used in the 1$D$ case, and the spectral diffusion length $\Lambda \sim [P(k_B T) k_B T]^{-1/3}$ ($P(E) \propto E^\mu$), relevant for a nonuniform distribution of defect energies, has been used. By accounting for the energy dependence of the DDOS measured by Skacel {\it et al.} \cite{Skacel15}, this theory \cite{Faoro15} correctly predicts the observed low-temperature enhancement of the $1/f$-noise observed in superconducting resonators \cite{Burnett14}. 

Similarly, the scaling of $S_\nu(\omega)$ at high field intensity is given by 
\begin{equation}
\label{}
S_\nu(\omega) \propto \frac{1}{\omega \sqrt{\mathcal{P}}}
 \left\{
\begin{array}{ll}
      \sqrt{\frac{\rho}{T_1}} T^{-(1+\mu)/2}            & D = 3  \\
     
      \sqrt{\frac{\rho_2}{T_1}}  T^{-(1+\mu)/3}     & D = 2 \\
     
      1                        & D = 1, \\
\end{array} 
\right. 
\end{equation} 
indicating that $S_\nu(\omega)$ is enhanced with lower temperature, and is suppressed as the system dimension is lowered. These results show that the temperature scaling of noise generated by strongly interacting defects has a unique dimension-dependent fingerprint, and that noise could be dramatically reshaped, through its dependence on $T_1$, in systems possessing a non-trivial phonon DOS.  

This concludes our discussion of defect-induced noise. In the following sections we explore linear and nonlinear absorption of EM and acoustic waves mediated by TLSs. 

\subsection{Defect-induced dissipation in mesoscale systems} 
Defects contribute a large source of dissipation in a number of mesoscopic optomechanical \cite{Anetsberger08,Arcizet09,Park09,Riviere11}, quantum information \cite{Simmonds04,Martinis05,Gao07,O'Connell08}, NEMS, and MEMS \cite{Zolfagharkhani05,Regal08,Hoehne10,Galliou11,Goryachev12,Galliou13,Goryachev13,Goryachev13c,Faust14} devices. As these systems push to ever-smaller sizes, changes in defect dynamics, the dispersion of acoustic modes, and the phonon DOS transform the character of defect-induced dissipation. In this section we investigate this transformation by showing how resonant and relaxation absorption, the two processes by which TLSs dissipate EM and acoustic waves, are determined by extrinsic system properties.  
\subsubsection{Geometric, dispersive, and Purcell enhancement of the nonlinear properties of resonant absorption}
Dissipation occurs via resonant absorption when a ground state defect absorbs a phonon or photon with energy matching its gap and then spontaneously reradiates in a random direction. Alternatively, amplification occurs when a phonon or photon incident on an excited defect elicits a decay via stimulated emission. Hence, resonant absorption scales with the difference of probabilities for a resonant defect to be in the ground vs. the excited state $p_g(E)-p_e(E)$.

When the EM and acoustic fields are weak, the defects remain in thermal equilibrium, and $p_g(E)-p_e(E) = \tanh(E/2 k_B T)$ for a defect at temperature $T$. In this limit, the dissipation rate for phonons (top) and photons (bottom) can be computed with Fermi's golden rule, giving the inverse quality factor (i.e. loss tangent) 
\begin{equation}
\label{resAbs-lowInt}
 \left.
\begin{array}{r}
    \frac{1}{Q^{\rm ac.}_{{\rm res},{\bf q}}}      \\
    \frac{1}{Q^{\rm em}_{{\rm res},{\bf q}}}     
\end{array} \!\!
\right\} =  \pi P_D(\hbar \Omega_{\bf q}) \left\{
\begin{array}{cc}
   \!\!\! \sum_\eta  \frac{ \gamma^2_\eta e_{{\bf q}\eta}}{ \rho_D v_\eta^2}  \!\!\! \\
    \frac{|{\bf d}|^2}{3 \varepsilon_0 \varepsilon_{\bf q}^{\rm eff}} \\
\end{array}  
\right\}  \tanh \left( \frac{\hbar \Omega_{\bf q}}{2 k_B T} \right),
\end{equation}
where a uniform density of defect positions and orientations has been assumed. This result shows a general characteristic of resonant absorption; namely, it saturates when $p_g(E)- p_e(E) \approx 0$, in this case at high temperatures.
In addition, Eq. \ref{resAbs-lowInt}, valid for confined fields, is reminiscent of the prediction made by the standard TSM, and only gives a small numerical correction to the dissipation for systems differing from 3$D$ bulk. However, this fortunate correspondence breaks down at high acoustic or EM intensities. 

As the intensity of the acoustic or EM field is raised, a point will be reached where the mean-free-time between defect-phonon or defect-photon interactions is equal to the defect's upper state lifetime. This critical intensity $J_c$ is met when the power incident on a defect is roughly one quanta per excited state lifetime
$
J_c \sim \frac{\hbar \Omega_{\bf q}}{\sigma T_1}
$ 
where $\sigma$ is the cross section for the absorption of a phonon or photon by a ground state defect. The cross section $\sigma$ can be obtained from Fermi's golden rule, giving the following expression for the acoustic $J^{\rm ac.}_c$ and EM $J^{\rm em}_c$ critical intensities 
\begin{equation}
\label{JC}
\left.
\begin{array}{r}
   J^{\rm ac.}_c     \\
    J^{\rm em}_c    
\end{array} \!\!
\right\}
\sim
\left\{ \!\!
\begin{array}{c}
    \frac{ \rho v^3 }{ \gamma^2 }      \\
     \frac{3  \varepsilon_0 \sqrt{\varepsilon} c}{ |{\bf d}|^2 }   
\end{array} \!\!
\right\} \frac{\hbar^2  }{2  T_1 T_2}.
\end{equation} 
 
At intensities exceeding $J_c$ incident phonons or photons begin to probe an excited defect before it returns to the ground state, allowing the defect to decay through stimulated emission and in turn to amplify the phonon or photon beam. Hence, at high-intensities absorption and amplification compensate one another, and resonant absorption is saturated. Equation \ref{JC} shows that this saturation scale is set by the defect dynamics, and therefore, the nonlinearity of resonant absorption is shaped by the extrinsic properties of the system.

In the high intensity regime, perturbation theory is no longer adequate to describe acoustic and EM dissipation, and the Bloch equations of the coupled system must be employed to describe resonant absorption (see Appendix \ref{Resonant-Appendix}).   
For an idealized $D$-dimensional system resonant absorption for plane-wave acoustic (top) or EM (bottom) modes of angular frequency $\Omega_{\bf q}$ and polarization $\eta$ is characterized by the inverse quality factor  
\begin{equation}
\label{resAbs-PW}
\left.
\begin{array}{r}
    \frac{1}{Q^{\rm ac.}_{{\rm res},{\bf q}}}      \\
    \frac{1}{Q^{\rm em}_{{\rm res},{\bf q}}}     
\end{array} \!\!
\right\} =   \frac{ \overline{P}_{D}(\hbar \Omega_{\bf q})}{4} \int d \varphi 
\left\{
\begin{array}{ll}
    \! \! \!    
\frac{\gamma^2_\eta(\hat{n})}{
\rho v_\eta^2}
  \! \! \!\! \\
     \! \! \!   \frac{d^2_\eta(\hat{n})}{3\varepsilon_0 \varepsilon}  \! \! \!\!
\end{array}  \right\} \frac{\tanh \left( \frac{\hbar \Omega_{\bf q}}{2 k_B T} \right)}{\sqrt{1+\frac{J}{J_{c}(\hat{n})}}} 
\end{equation}
where $J$ and $J_c(\hat{n})$ is the intensity and orientation-dependent critical intensity for the acoustic or EM field, i.e. the acoustic intensity is $J^{\rm ac.} \equiv  (\hbar \rho v_\eta^3/\Omega_{\bf q})|\underline{\xi}_{\bf q} \beta_{\bf q}|^2$ and the EM intensity is $J^{\rm em} =  \varepsilon_0 \sqrt{\varepsilon} c \hbar \Omega_{\bf q}|{\bf E}_{\bf q} \alpha_{\bf q}|^2$ ($\beta_{\bf q}$ and $\alpha_{\bf q}$ being the amplitude for the {\bf q}th phonon and photon mode, respectively). 
The orientation-dependent deformation potential for coupling to $\eta$-polarized acoustic plane waves is given by
\begin{align}
\label{DefPots}
 \gamma_\ell(\hat{n})   &  =  \tilde{\gamma}(1-2\zeta \sin^2 \theta)   \\
 \gamma_{t,1}(\hat{n}) &  =  \tilde{\gamma} (2\zeta \sin \theta \cos \theta \cos \phi) \nonumber \\
 \gamma_{t,2}(\hat{n})  &  =  \tilde{\gamma} (2\zeta \sin \theta \cos \theta \sin \phi),\nonumber
\end{align}
and $d_\eta(\hat{n})$ is given by $ \sqrt{3} |{\bf d}|\cos \theta$ where the ${z}$-axis of the dipole orientation coordinate system has been chosen to align with electric field. $\int d \varphi$ is an integral over solid angle, and $\overline{P}_{D}(\hbar \Omega_{\bf q})$ is the spatial average of the DDOS. The spatial averaging is simplified for idealized bulk systems because of the spatial-independence of $T_1$ and the modulus of the acoustic and EM spatial eigenfunctions. 

The exact form of $J_c$ matches well with the result anticipated from the basic timescale arguments that led to Eq. \eqref{JC}
\begin{equation}
\label{JCexact}
\left.
\begin{array}{r}
   J^{\rm ac.}_c(\hat{n})     \\
    J^{\rm em}_c(\hat{n})    
\end{array} \!\!
\right\}
 = \left\{ \!\!
\begin{array}{c}
     \frac{\rho_D v_\eta^3}{ \gamma_\eta^2(\hat{n}) }   \\
        \frac{3 \varepsilon_0 \sqrt{\varepsilon} c}{ d^2_\eta(\hat{n})} 
\end{array} \!\!
\right\}  \frac{\hbar^2}{2 T_{1,{\rm min}}T_2},
\end{equation}
but we emphasize that the $D$-dimensional forms of $T_{1,{\rm min}}$ and $T_2$ must be used. The standard TSM prediction for the damping factor $1/Q_{{\rm res},{\bf q}}$ at high-intensity can be obtained from Eq. \eqref{resAbs-PW} by taking $\gamma_\eta(\hat{n})\to \gamma_\eta$ and $d_\eta(\hat{n}) \to d_\eta$ in the critical intensity of Eq. \eqref{JCexact}. 

Equation \ref{resAbs-PW} has a similar functional form to resonant absorption in the strong field regime given by the standard tunneling state model \cite{Phillips87}. However, striking differences arise from the dependence of the critical intensity on $T_1$ and $T_2$, and hence the distinct physics of resonant absorption in idealized bulk systems is largely characterized by a dimensional modification of the magnitude, temperature and frequency dependence of $J_c$. 
 
\begin{figure}[h]
\begin{center}
\includegraphics[width=0.45\textwidth]{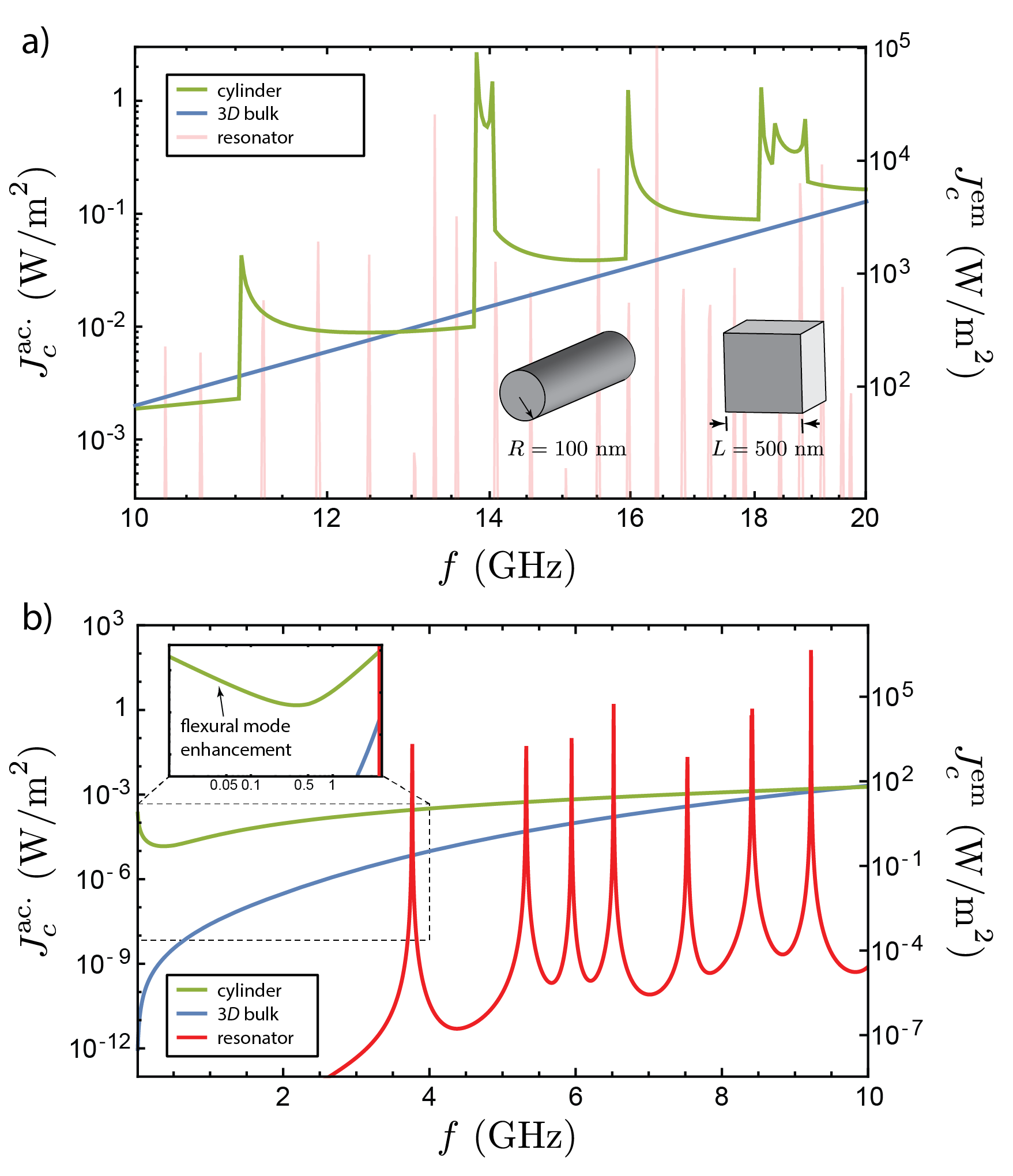}
\caption{Critical intensity at 10 mK in a silica cylinder (green), resonator (red), and 3$D$ bulk as a function of frequency computed using Eqs. \ref{T1-ave}, \ref{JCexactWG}, and \ref{T2-tempScaling}. The cylinder exhibits enhancements of the critical intensity at van-Hove singularities a) and at low frequencies b) where $T_1^{-1}$ is dominated by emission into flexural modes. The critical intensity is Purcell enhanced in the resonator. Deformation potential and sound velocity for longitudinal waves and $|{\bf d}| = 1.3$ Debye and $\varepsilon = 2.08$  were used in Eq. \eqref{JC}.}
\label{criticalIntensity}
\end{center}
\end{figure}

The behavior of resonant absorption is nontrivial in mesoscale systems possessing flexural, slow-group velocity, or standing wave modes. We show that the critical intensity is enhanced at low frequencies due to flexural modes, sharply increases near van-Hove singularities in the phonon DOS, and is Purcell enhanced in resonators (see Figs. \ref{criticalIntensity}-\ref{Q-Results-Resonator}). Moreover, the spatial dependence of the energy density may have poor overlap in systems with an anisotropic DDOS, such as those constructed from crystalline media where defects are concentrated on surfaces and at interfaces. In such mesoscale systems resonant acoustic absorption is characterized by the quality factor  
\begin{equation}
\label{resAbs-Gen}
\frac{1}{Q^{\rm ac.}_{{\rm res},{\bf q}}} =  \left<  \frac{\pi V_D  P_D}{ \Omega_{\bf q}^2} 
\frac{  |\boldsymbol{\gamma}:{\underline{\xi}_{\bf q}}|^2  \tanh \left( \frac{\hbar \Omega_{\bf q}}{2 k_B T} \right)}
{\sqrt{1+\frac{J_{ave}}{J^{ac.}_c(\hat{n},{\bf r}) }}}
\right>_V.
\end{equation}
 For arbitrary guided traveling waves the critical intensity is given by  
\begin{align}
\label{JCexactWG}
J_c^{ac.}(\hat{n},{\bf r}) = & \frac{(\hbar \Omega_{\bf q})^2 v_g}{2 T_{1,{\rm min}} T_2 |\boldsymbol{\gamma}:{\underline{\xi}_{\bf q}}({\bf r})|^2 V} 
\end{align}
which depends on defect orientation and position. In contrast to idealized bulk systems $J_c^{ac.}(\hat{n},{\bf r})$ scales with the group velocity $v_g$ of the driven mode, suggesting that the nonlinearity of the system may be engineerable. For resonators $J_{ave}/J^{ac.}_c(\hat{n},{\bf r})$ should be taken to $\mathcal{E}_{ave}/\mathcal{E}^{ac.}_c(\hat{n},{\bf r})$, the ratio of average mode energy density to the critical mode energy density, in Eq. \eqref{resAbs-Gen} defined by $\mathcal{E}_{ave} V = \hbar \Omega_{\bf q} |\beta_{\bf q}|^2$ and $\mathcal{E}^{ac.}_c(\hat{n},{\bf r}) = J_c^{ac.}(\hat{n},{\bf r})/v_g$. The loss tangent for the EM field in an arbitrary structure can be obtained from Eq. \eqref{resAbs-Gen} by taking $|\boldsymbol{\gamma}:{\underline{\xi}_{\bf q}}| \to \Omega_{\bf q} |{\bf d}\cdot {\bf E}_{\bf q}|$. 

To illustrate the transformation of the nonlinear behavior of resonant absorption in mesoscale systems, the critical intensity for a silica microwire, resonator, and 3$D$ bulk system are compared as a function of frequency in Fig. \ref{criticalIntensity}. For simplicity, the orientation- and spatial-averaged defect decay rate is used in Eq. \eqref{JCexactWG}, $|\boldsymbol{\gamma}:{\underline{\xi}_{\bf q}}|^2$ is replaced with $\langle |\boldsymbol{\gamma}:{\underline{\xi}_{\bf q}}|^2 \rangle_V$, and the resonator is modeled by using periodic boundary conditions (which is why an intensity can be defined). The wire radius and temperature are chosen so that $T_2$ is dimensionally reduced. Above $\Omega_{co}$ the critical intensity for resonant absorption in the cylinder exhibits sharp enhancements at van-Hove singularities in the phonon DOS (Fig. \ref{criticalIntensity} a), and the critical intensity in the resonator is Purcell enhanced at resonator mode frequencies. At low frequencies the critical intensity in the microwire is enhanced by dispersive flexural modes (Fig. \ref{criticalIntensity} b). 

\subsubsection{Geometric, dispersive, and DOS transformations of relaxation absorption}
In this section we discuss the transformation of relaxation absorption by reduced dimensionality, phonon dispersion, and confinement. Relaxation absorption is a non-resonant source of dissipation that occurs when phonons or photons modulate TLS energy levels. In this process, defects are driven in and out of thermal equilibrium with their environment leading them to absorb energy from phonons or photons and release it to the environment in an irreversible fashion. Unlike resonant absorption, relaxation absorption is not saturable, and thus it sets the minimum level of dissipation that can be achieved in a system containing defects. We show that this form of dissipation is enhanced in many mesoscale systems by dispersion and confinement, but in contrast, it is exponentially suppressed in resonators at low temperatures.    

We begin this discussion of relaxation absorption by stating the result for the quality factor $Q_{{\rm rel},{\bf q}}$ for an arbitrary system (see Appendix \ref{Relaxation-Appendix})
\begin{align}
\label{relAbs}
\left.
\begin{array}{r}
    \frac{1}{Q^{\rm ac.}_{{\rm rel},{\bf q}}}      \\
    \frac{1}{Q^{\rm em}_{{\rm rel},{\bf q}}}     
\end{array}
\right\}
& = & 
  \frac{V_D}{\Omega_{\bf q}^2 k_B T }   \int d\Delta d\Delta_0 \frac{\Delta^2}{\Delta_0 E^2} \sech^2\left(\frac{E}{2 k_B T}\right)\nonumber\\
&   &
  \times \left< P_D(E) \left\{
  \begin{array}{c}
       | \boldsymbol{\gamma}:{\underline{\xi}_{\bf q}}|^2    \\
          \Omega_{\bf q}^2|  {\bf d}\cdot {\bf E}_{\bf q}|^2
\end{array}
\right\}
    \frac{\Omega_{\bf q} T_1}{1+\Omega_{\bf q}^2 T_1^2}  
  \right>_V
\end{align}
which accounts for the mode structure of the field, the position and orientation of all defects, and modifications of the phonon DOS. Notice that the relative contribution of defects of different energies to this process is determined by the factor $\sech^2\left(E/2 k_B T\right)$, and therefore the contribution from TLSs, and also phonons, with $E > k_B T$ is exponentially suppressed. Hence, this process is dimensionally reduced when the frequency of thermal phonons, $\omega_{\rm th} \ (\sim 208 \ {\rm MHz} (T/10 \ {\rm mK})$), is much less than a structure's cutoff frequency $\Omega_{co}$ (e.g. see Figs. \ref{dimensionalRedux} \& \ref{dispRelation}).    

To understand the qualitative behavior of relaxation absorption as the system dimension is lowered we consider
idealized bulk $D$-dimensional systems. To gain more insight from Eq. \eqref{relAbs} we consider ranges of parameters where $Q_{{\rm rel},{\bf q}}$ can be approximated. When $\Omega_{\bf q} T_{1,\rm min}(E < k_B T) \gg 1$
the integrand of Eq. \eqref{relAbs} can be Taylor expanded for small $1/(\Omega_{\bf q} T_1)$ and the integrals can be done analytically. This calculation results in the asymptotic form $Q_{{\rm rel},{\bf q}}$ for $\eta$-polarized phonons and photons of frequency $\Omega_{\bf q}$ given by
\begin{align}
\label{relAbs-HF-Bulk}
\left.
\begin{array}{r}
    \frac{1}{Q^{\rm ac.}_{{\rm rel},{\bf q}}}      \\
    \frac{1}{Q^{\rm em}_{{\rm rel},{\bf q}}}     
\end{array}
\right\} \approx &  
  \frac{P_D(k_B T)}{3 \Omega_{\bf q} \rho_D \hbar^{D+1}} \frac{\pi S_{D-1} 2^{D+1+\mu}}{(2\pi)^D}  \nonumber \\
  & \times
   \sum_{\eta'}
   \left\{ 
   \begin{array}{l}
        \frac{[\gamma^4]_{\eta \eta'}}{\rho_D v_\eta^2} \\    
          \frac{[d^2 \gamma^2]_{\eta \eta'}}{\varepsilon_0 \varepsilon}   
\end{array}
\right\}
  \frac{\mathcal{I}_{D+\mu}}{v_{\eta'}^{D+2}} (k_B T)^{D} 
\end{align} 
where $[\gamma^4]_{\eta \eta'} \equiv (4\pi)^{-1}\int d\varphi \ \gamma_\eta^2(\hat{n})\gamma^2_{\eta'}(\hat{n})$, $[d^2 \gamma^2]_{\eta \eta'} \equiv (4\pi)^{-1}\int d\varphi \ d_\eta^2(\hat{n}')\gamma^2_{\eta'}(\hat{n}) = |{\bf d}|^2 \gamma^2_{\eta'}/3$. Here we do not assume that a defect's electric dipole and deformation potential are parallel, i.e. $\hat{n}'\neq \hat{n}$ \cite{Footnote-2}, and $P_D(E) \propto E^\mu$. The familiar result from the standard TSM is obtained by ignoring the angle dependence of the deformation potential, taking $D = 3$, and assuming a constant DDOS, i.e. $\mu = 0$ \cite{Phillips87}. Up to a pure numerical prefactor, Eq. \eqref{relAbs-HF-Bulk} shows that relaxation absorption is geometrically enhanced  by a factor $(L/\lambda_{th})^D$ in lower dimensional bulk systems. We also point out that the results of Eq. \eqref{relAbs-HF-Bulk} agree with recent measurements of dissipation, attributed to phonon-mediated relaxation, of quasi-1$D$ NEMS oscillators exhibiting a linear temperature scaling \cite{Hoehne10}. 

In the low-frequency limit where ${\Omega_{\bf q} T_{\rm 1,min}(E = k_B T) \ll 1}$ relaxation absorption reduces to a universal value that is independent of the system dimension. This can be seen by converting the integration in Eq. \eqref{relAbs} to `polar' coordinates, i.e. $(\Delta,\Delta_0) \to (E,\phi)$, the $\phi$ integral can be done directly. Subsequently, a Taylor expansion in small $\Omega_{\bf q} T_{\rm 1,min}$ may be performed resulting in 
\begin{align}
\label{relAbs-Universal}
\left.
\begin{array}{r}
    \frac{1}{Q^{\rm ac.}_{{\rm rel},{\bf q}}}      \\
    \frac{1}{Q^{\rm em}_{{\rm rel},{\bf q}}}     
\end{array}
\right\} & =  
  \pi P_D(k_B T)  2^{\mu-1}c_\mu
  \left\{ \begin{array}{l}
       \frac{\gamma_\eta^2}{\rho_D v_\eta^2}   \\
         \frac{|{\bf d}|^2}{3 \varepsilon_0 \varepsilon^{\rm eff}_{\bf q}}
\end{array}
\right\}.
  \end{align} 
Thus, in the low-frequency limit, the temperature scaling of $1/Q_{{\rm rel},{\bf q}}$ is given by $P(k_B T) \propto T^\mu$ which provides an indirect window on the energy dependence of the DDOS. Also, it is interesting to note that Eq. \eqref{relAbs-Universal} with the measured value of $\mu \approx 0.3$ leads to a low-temperature scaling of the mechanical dissipation in agreement with observations in quartz BAW resonators, and a variety of NEMS and MEMS that operate in the low-frequency limit \cite{Zolfagharkhani05,Galliou13}. However, such a scaling can be explained by relaxation absorption associated with overdamped flexural modes in the high-temperature limit \cite{Seoanez07a}. These results point to an interesting direction for further study.   

Acoustic waveguides often support dispersive flexural modes without cutoff. In such systems the reduced dimensional behavior of relaxation absorption contrasts with the results of Eq. \eqref{relAbs-HF-Bulk}. In waveguides, the dispersion of each phonon branch can lead to dramatic changes in the DOS of the phonon bath, and in turn modify the temperature scaling and magnitude of acoustic and EM dissipation. To explore these effects with maximum simplicity, we compute $1/Q_{{\rm rel},{\bf q}}$ for waveguides in the high-frequency limit using the spatial and orientation averaged value of $T_1$ in Eq. \eqref{relAbs}, and consider systems where $(\gamma_\ell/v_\ell)^2 \approx (\gamma_t/v_t)^2$. With these approximations and using Eq. \eqref{T1-ave} we find 
 \begin{align}
\label{relAbs-1Dwg-Simp}
\left.
\begin{array}{r}
    \frac{1}{Q^{\rm ac.}_{{\rm rel},{\bf q}}}      \\
    \frac{1}{Q^{\rm em}_{{\rm rel},{\bf q}}}     
\end{array} \!\!
\right\} \approx & 
  \int_0^\infty d E \frac{2\pi P(E) E}{3\hbar^2 \Omega_{\bf q} \rho k_B T V} g(E/\hbar) \frac{\gamma_\ell^2}{v_\ell^2} 
 \nonumber \\
 &  \quad \quad  \times  \left\{\!\! \begin{array}{c }
       \gamma_\ell^2/\rho v_\ell^2    \\
       |{\bf d}|^2/3 \varepsilon_0 \varepsilon^{\rm eff}_{\bf q} 
\end{array} \!\!
\right\} {\rm csch}\left(\frac{E}{k_B T}\right)   
\end{align}
where the identity $2 {\rm csch} (x) = \sech^2(x/2) \coth(x/2)$ has been used. The phonon DOS $g(\Omega)$ for 1$D$ and 2$D$ systems is given by Eqs. \ref{1D-DOS} and \ref{2D-DOS}.

When $k_B T \ll \hbar\Omega_{co}$, the behavior of waveguides differs from the idealized bulk systems. To see this consider a cylindrical waveguide of radius $R$ where only the four cylinder modes without cutoff contribute to Eq. \eqref{relAbs} (Fig. \ref{dispRelation}). The effect of the compressional and torsional mode is accurately predicted by Eq. \eqref{relAbs-HF-Bulk}, but in contrast, the magnitude and temperature scaling of relaxation absorption from flexural modes differs substantially from bulk systems. To see this consider small wavevectors, i.e. $R q \ll1$, where the dispersion relation for flexural modes in a cylinder is given by $\Omega \approx v_B R q^2/2$ \cite{RoyerBook}. In this limit the inverse quality factor resulting from the flexural mode contribution to $1/Q_{{\rm rel},{\bf q}}$ is given by
 \begin{align}
\label{relAbs-1Dwg-Flex}
\left.
\begin{array}{r}
    \frac{1}{Q^{\rm ac.}_{{\rm rel},{\bf q}}}      \\
    \frac{1}{Q^{\rm em}_{{\rm rel},{\bf q}}}     
\end{array} \!\!\!
\right\}_{\rm flex.,1D}  \approx & 
  \frac{2^{1+\mu}}{3\Omega_{\bf q} \rho  A}  \frac{\gamma_\ell^2}{v_\ell^2} \frac{P(k_B T)}{\sqrt{v_B R}} 
  \nonumber \\ & \times 
  \left\{ \!\!\! \begin{array}{c }
       \frac{\gamma_\ell^2}{\rho v_\ell^2}    \\
       \frac{|{\bf d}|^2}{3 \varepsilon_0 \varepsilon^{\rm eff}_{\bf q}} 
\end{array} \!\!\! 
\right\} 
   \frac{(k_B T)^{1/2}}{\hbar^{3/2}} \mathcal{I}_{1/2+\mu}
\end{align}
(recall that $\mathcal{I}_{m}$ is defined following Eq. \eqref{rel-Abs-HF-noise}).
The result above shows that relative magnitude of Eq. \eqref{relAbs-1Dwg-Flex} to the quality factor for an idealized bulk $1D$ system is enhanced by $\sqrt{\lambda_{th}/R}$ ($\gg$ 1 when $\hbar \Omega_{co} \gg k_B T$), showing that flexural modes dominate relaxation absorption in waveguide systems. Losses in such systems scale as $P(k_B T)T^{1/2} \propto T^{1/2+\mu}$ when resonant absorption is negligible, and may be an alternative explanation for the temperature dependence of mechanical dissipation observed in nano-beams \cite{Hoehne10}.    

Flexural modes in 2$D$ waveguides also lead to enhancement of the acoustic decay over the bulk $2D$ result above. 
A similar analysis as that performed for the cylinder above gives the relaxation absorption related quality factor arising from flexural modes in a planar system
\begin{align}
\label{relAbs-2Dwg-Flex}
\left.
\begin{array}{r}
    \frac{1}{Q^{\rm ac.}_{{\rm rel},{\bf q}}}      \\
    \frac{1}{Q^{\rm em}_{{\rm rel},{\bf q}}}     
\end{array} \!\!\!
\right\}_{\! \rm flex.,2D}  \approx & 
  \frac{2^{1+\mu}}{\sqrt{3}\Omega_{\bf q} \rho L}  \frac{\gamma_\ell^2}{v_\ell^2} \frac{P(k_B T)}{v_{pl} L} 
    \left\{ \!\!\! \begin{array}{c }
       \frac{\gamma_\ell^2}{\rho v_\ell^2}    \\
       \frac{|{\bf d}|^2}{3 \varepsilon_0 \varepsilon^{\rm eff}_{\bf q}} 
\end{array} \!\! \!
\right\}   \frac{k_B T}{\hbar^2} \mathcal{I}_{1+\mu} 
\end{align}
where the dispersion relation for the fundamental flexural mode $\Omega\approx \frac{1}{2\sqrt{3}} v_{pl} L k^2$, valid for $L k \ll 1$, has been used \cite{RoyerBook}. A comparison of Eq. \eqref{relAbs-2Dwg-Flex} to Eq. \eqref{relAbs-HF-Bulk} shows that flexural modes are the dominant source of relaxation absorption in 2$D$ structures at low temperatures. 

Now we analyze relaxation absorption in resonant acoustic cavities. For cavities the decay rate $T^{-1}_1$ is Purcell enhanced, and so care must be taken when assessing the asymptotic limits of $Q_{{\rm rel},{\bf q}}$. In high finesse cavities it may be possible that ${\Omega_{\bf q} T_1 \gg 1}$ for non-resonant defects and ${\Omega_{\bf q} T_1 \ll 1}$ for resonant defects. 
However, in certain ranges of frequencies and temperatures, the inequality ${\Omega_{\bf q} T_1 \gg 1}$, is satisfied for all energies contributing to the integral in Eq. \eqref{relAbs}. In this limit, the peaked nature of the TLS-decay rate inside the integrand samples energies matching phonon resonances and leads to the quality factor given by 
\begin{align}
\label{relAbs-Cavity-2}
\left.
\begin{array}{r}
    \frac{1}{Q^{\rm ac.}_{{\rm rel},{\bf q}}}      \\
    \frac{1}{Q^{\rm em}_{{\rm rel},{\bf q}}}     
\end{array}
\right\} \approx & 
  \frac{2\pi }{3 \Omega_{\bf q} V_D k_B T}\frac{\gamma_\ell^2}{v_\ell^2}  \sum_{\bf q'}  \frac{ P_D(\hbar \Omega_{\bf q'}) \Omega_{\bf q'}
   }{ \sinh \frac{\hbar \Omega_{\bf q'}}{k_B T} } 
    \left\{ \begin{array}{c }
     \!\!\!   \frac{\gamma_\ell^2}{\rho v_\ell^2}   \!\!\!   \\
      \!\!\!   \frac{|{\bf d}|^2}{3 \varepsilon_0 \varepsilon^{\rm eff}_{\bf q}}  \!\!\! 
\end{array} 
\right\}.
\end{align} 
In the limit where the mode volume becomes large the resonator eigenfrequencies become dense and Eq. \eqref{relAbs-Cavity-2} reduces to Eq. \eqref{relAbs-HF-Bulk}. In contrast, for small mode volumes the phonon spectra is gapped, and as was seen for TLS-induced noise in resonators, relaxation absorption too is exponentially suppressed for low temperatures (i.e. $k_B T < \hbar \Omega_1$) (Fig. \ref{Q-Results-Resonator}). 
 
\subsection{Estimations of TLS-induced dissipation in mesoscale waveguides and resonators}
To illustrate the contrasting behavior of defect-induced dissipation in mesoscale systems we compare the total the acoustic quality factor and EM loss tangent, given by $Q^{-1}_{\bf q} = Q^{-1}_{{\rm res},{\bf q}} + Q^{-1}_{{\rm rel},{\bf q}}$, for waveguides, resonators, and a 3$D$ bulk. The total quality factor is computed from Eqs. \ref{resAbs-Gen} and \ref{relAbs} and the expressions for $T_1$ and $T_2$. For simplicity, the spatial and orientation averaged defect decay rate (see Eqs. \ref{T1-guided wave-1D}-\ref{T1,cav}), including the effect of dispersive higher order modes and Purcell enhancement, is used to compute the $J_c$ and $Q_{{\rm rel},{\bf q}}$ in these examples. The defect dephasing rate is given by $T_2^{-1} = T_1^{-1}/2+T_{2}^{-1'}$ where Eqs. \ref{T2-tempScaling} and the convention discussed at the end of Sec. IV C are used. 

\subsubsection{Dissipation in nanoscale waveguides}
\begin{figure}
\begin{center}
\includegraphics[width=0.45\textwidth]{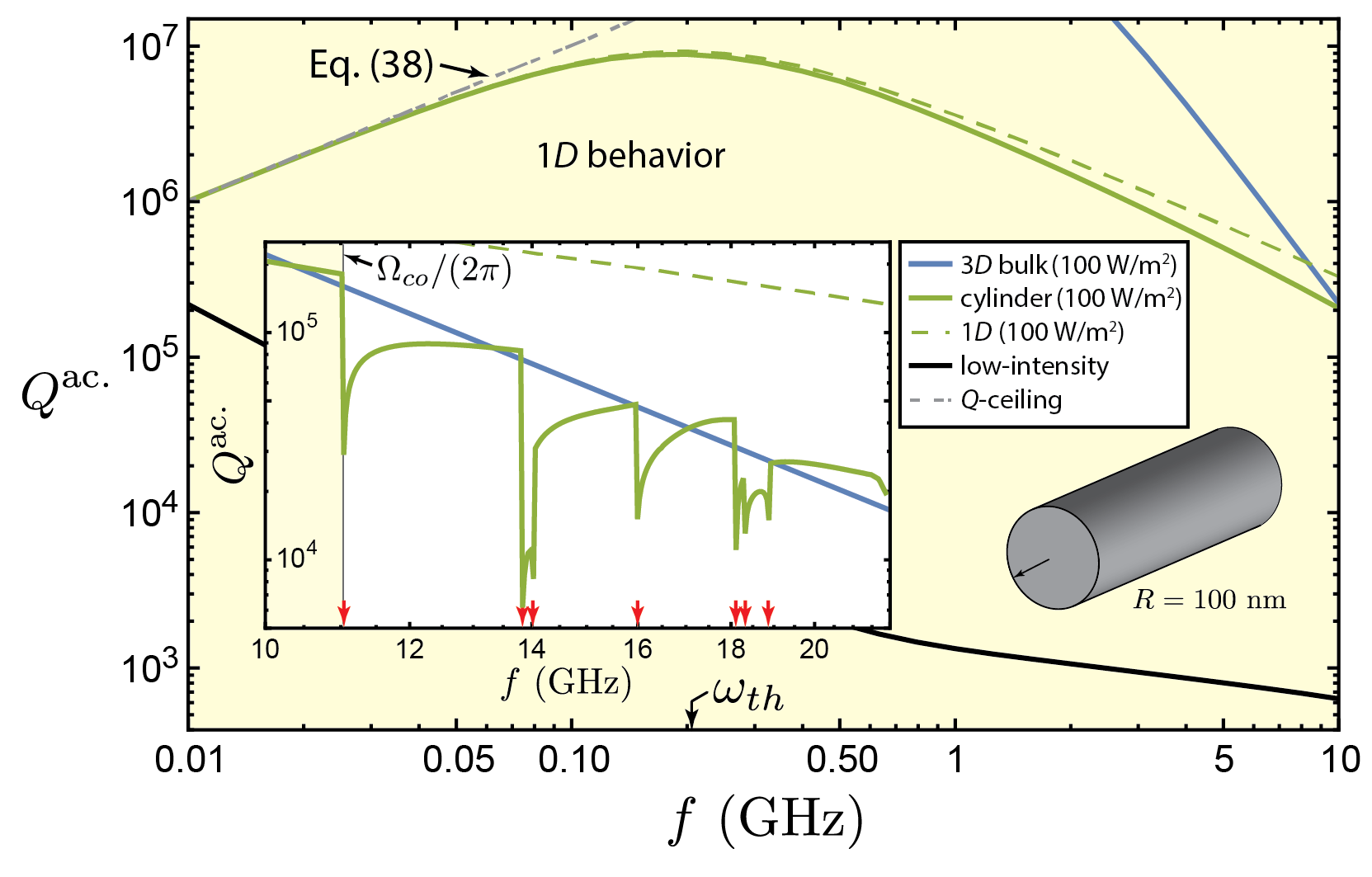}
\caption{Acoustic quality factor for the fundamental axial-radial mode of a $100$ nm radius silica wire as a function of frequency (green) with 100 ${\rm W/m}^2$ intensity. The wire temperature is 10 mK. For comparison, $Q$-factor for quasi-1$D$ bulk (green dashed) and a 3$D$ bulk (blue) with the same parameters are displayed, as well as the low- and high-intensity limits, respectively (solid black) and (gray dashed). Inset: $Q$-factor for frequencies above cutoff showing large changes near van-Hove singularities in the phonon DOS (red-arrows).}
\label{Q-Results-Waveguide}
\end{center}
\end{figure}

Figure \ref{Q-Results-Waveguide} illustrates the contrasting behavior of the acoustic quality factor of the fundamental axial-radial mode of a cylindrical nanowire and a longitudinal wave of an idealized $3D$ bulk as a function of frequency. Both excitations are chosen to have an intensity of $100$ W/m$^2$, and both systems are set at a temperature of 10 mK. The microwire system is chosen for its simplicity; its mode functions can be obtained analytically, yet it exhibits all of the unique behaviors of waveguides.  These behaviors include dispersive flexural modes, and van-Hove singularities in the phonon DOS.  For the chosen microwire radius (100 nm) and temperature, the thermal frequency is far below cutoff (i.e. $k_B T \ll \hbar \Omega_{co}$, see Fig. \ref{dimensionalRedux}). Therefore relaxation absorption is dimensionally reduced, with a magnitude determined by the four acoustic modes without cutoff, but dominated by flexural modes.   

Figure \ref{Q-Results-Waveguide} shows that the quality factor of the axial-radial mode in the cylinder (green) is much smaller than the 3$D$ bulk system (blue) for frequencies below $\Omega_{co}$. This occurs because the critical intensity is geometrically and dispersively enhanced in the microwire (see Fig. \ref{criticalIntensity}), and because relaxation absorption is geometrically enhanced by flexural modes Eq. \eqref{relAbs-1Dwg-Flex}. The reduced dimensional theory for the cylinder computed using Eq. \eqref{T1-cyl} (green dashed line of Fig. \ref{Q-Results-Waveguide}) differs from the exact calculation with higher frequency due to higher order dispersion not accounted for in Eq. \eqref{T1-cyl}. Above $\Omega_{co}$ sharp discontinuities are observed in $Q^{\rm ac.}$ at frequencies where acoustic excitations with zero group velocity are supported (i.e. van-Hove singularities), contrasting markedly from bulk systems (see Fig. \ref{Q-Results-Waveguide}). The dissipation is dominated by resonant absorption at low intensities, and the $Q$ of all systems converges to a nearly universal value determined by Eq. \eqref{resAbs-lowInt} (black line of Fig. \ref{Q-Results-Waveguide}).
While this example focuses on phononic dissipation, the EM loss tangent has a very similar character.

\subsubsection{Dissipation in acoustic resonators}
\begin{figure}
\begin{center}
\includegraphics[width=3.3in]{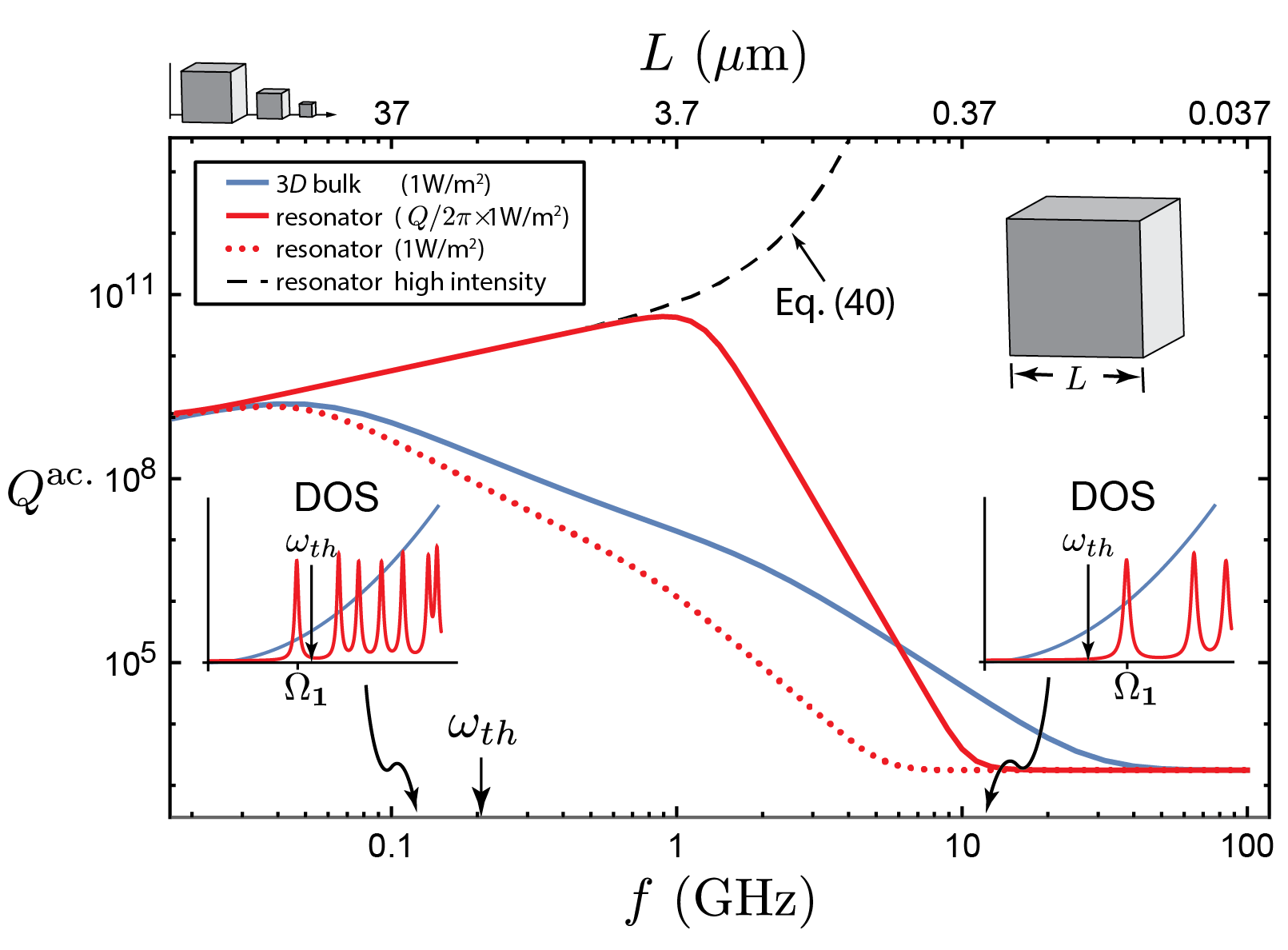}
\caption{Acoustic quality factor of the fundamental acoustic mode of a silica resonator with acoustic intensity of $1$ W/m$^2$ (red-dotted), and finesse-enhanced intensity $\frac{Q}{2\pi} 1$W/m$^2$ (red). The system temperature is taken to be 10 mK. The quality factor ceiling for the resonator, given by Eq. \eqref{relAbs-Cavity-2}, is shown as a black-dashed line. For comparison the result for a $3D$ bulk system with intensity 1W/m$^2$ is displayed (blue).}
\label{Q-Results-Resonator}
\end{center}
\end{figure}

The Purcell effect and the gapped phonon DOS lead to marked differences between
defect-induced dissipation in resonators and bulk systems. Such differences are displayed in Fig. \ref{Q-Results-Resonator}  which compares $Q^{\rm ac.}$ for the fundamental shear wave of a cubic silica resonator to the fundamental shear wave of a $3D$ bulk as a function of frequency. The resonator is defined using periodic boundary conditions and the frequency of the fundamental shear mode is continuously varied by scaling the resonator size. This example is amenable to a fully analytical treatment, and it has all of the salient features of a general resonator system.  

We have shown that the critical intensity for a resonator mode is Purcell enhanced (Fig. \ref{criticalIntensity}), and as a result $Q^{\rm ac.}$ is smaller in resonators (red-dotted) than in 3$D$ bulk systems (blue) for fixed circulating intensity (1 W/m$^2$) (Eq. \eqref{resAbs-Gen} \& Fig. \ref{Q-Results-Resonator}). However, the intra-cavity power is enhanced in resonators by energy storage. Therefore, we also compare $Q^{\rm ac.}$ for resonator and bulk systems with the same driving intensity (1 W/m$^2$), and including the effects of intra-cavity power enhancement. When this enhancement is accounted for, resonators (red) out-perform bulk systems (blue) over a broad range of frequencies.  

In resonators the quality-factor ceiling set by relaxation absorption is exponentially enhanced when at low temperatures($k_B T \ll \hbar \Omega_1$, see Eq. \ref{relAbs-Cavity-2}). This enhancement is illustrated by the black dashed line in Fig. \ref{Q-Results-Resonator}. For the example of Fig. \ref{Q-Results-Resonator}, with cubic geometry and plane wave normal modes, relaxation absorption in resonators and bulk systems is equivalent when $\Omega_1 \ll \omega_{th}$. However, when $\Omega_1 > \omega_{th}$ the resonator's thermo-mechanical motion is frozen out and $Q_{\rm rel}$ is enhanced (black-dashed line).

As a final remark we mention that the resonator is assumed to contain an ensemble of defects in these examples. However, if a uniform DDOS is assumed then fewer defects are contained in the system as its dimensions are scaled down. In such scaled down systems, with a small number of defects, Fig. \ref{Q-Results-Resonator} describes the average quality factor (the observed dissipation will fluctuate from sample-to-sample). 

\subsubsection{Low- and high-intensity limits of defect-induced dissipation}
\begin{figure}[ht]
\begin{center}
\includegraphics[width=0.45\textwidth]{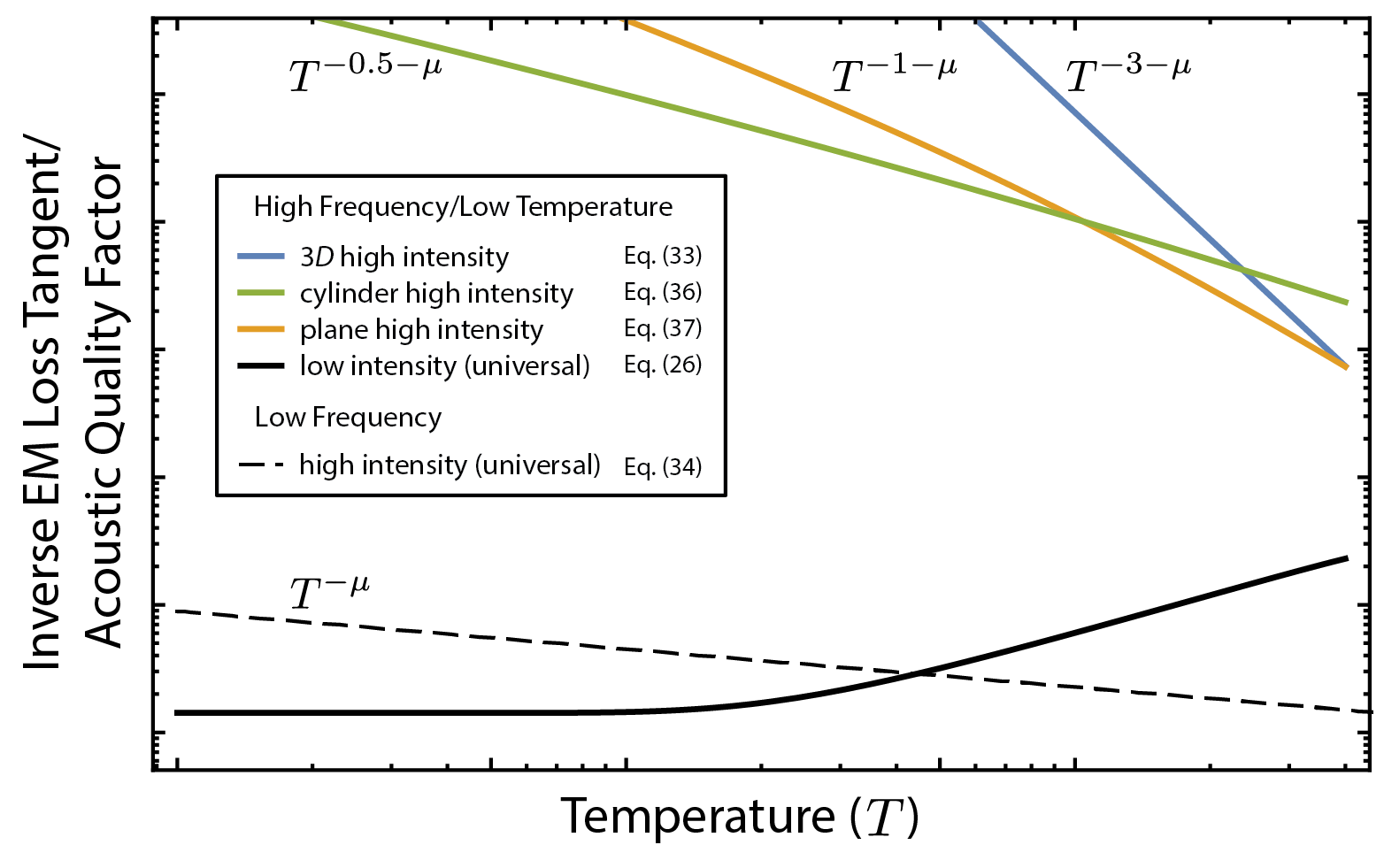}
\caption{Asymptotic limits of inverse EM and acoustic quality factor at high- and low-intensities. The parameter $\mu$ is set to 0.3.}
\label{Q-Results}
\end{center}
\end{figure}

Above, we saw that mesoscopic systems exhibit nontrivial saturation and dissipation characteristics determined by the details of the phonon DOS. However, in certain limits there are several striking universal trends, shown in Fig. \ref{Q-Results}, that are common to EM and acoustic dissipation. 

The dissipation is universal in two limits, surprisingly having nearly the same magnitude for EM and acoustic fields with the same mode volume. The first regime is reached at low intensity ($J\ll J_c$) and low-temperatures ($k_B T < \hbar \Omega$) where resonant absorption, described by Eq. \eqref{resAbs-lowInt}, dominates the dissipation (black). In this limit the temperature dependence of the dissipation is determined by the thermal population inversion of the defects that interact with the wave of interest. The second limit is reached at low-frequencies where relaxation absorption converges to a universal value given by Eq. \eqref{relAbs-Universal} (black-dashed). In this limit, the temperature dependence of this universal trend probes the energy dependence of the DDOS. 
 
When an arbitrary mode is driven to saturation the EM and acoustic dissipation approaches a universal (but system-dimension dependent) dissipation floor set by relaxation absorption. In Fig. \ref{Q-Results} we display results for systems where relaxation absorption dimensionally reduced ($\omega_{th} < \Omega_{co}$), and therefore the temperature dependence is determined by the system dimension, the energy dependence of the DDOS, and dispersive properties of the fundamental acoustic modes. In Fig. \ref{Q-Results} the high-intensity limit for the acoustic quality factor and inverse loss tangent in 1 (green), 2 (orange), and 3$D$ (blue) systems supporting flexural modes is shown. The low-temperature scaling of $T^{1/2+\mu}$, $T^{1+\mu}$, and $T^{3+\mu}$ for 1, 2, and 3$D$ respectively, serves as a powerful diagnostic measurement to survey the mechanical degrees of freedom and the DDOS that contributes to dissipation in a given system.  As a final note in this section, these results show that the unique properties of mesoscale systems are only visible at high intensities. 

\section{Discussion}
As an array of emerging nanoscale technologies progress to ever-smaller sizes the interplay of geometry, dispersion and density of states lead to radical modifications of the nature of defect-induced noise, dissipation, and nonlinearity.

We have shown that the nature of defect dynamics is determined by the interplay of confinement, TLS energy, defect concentration, and temperature. Namely, emission into slow group velocity, flexural, or resonator modes leads to a large (Purcell) enhancement of the decay rate, and when the separation between thermally activated defects exceeds one or more system dimension the behavior of spectral diffusion is transformed (see Figs. \ref{dimensionalRedux}, \ref{T1-cyl-fig} \& \ref{T1,cav}). As a result, the noise produced by defects is shaped by system geometry and is suppressed in systems constructed from high-quality acoustic resonators operating at low-temperatures (see Figs. \ref{PS-Single-Defect-Cavity}, \ref{Ensemble-Noise-Bulk} \& \ref{PS-Ensemble-Cavity}). In addition, the saturation scale for defect-induced dissipation (Fig. \ref{criticalIntensity}), and the dissipation floor at high-intensities is strongly modified by geometric, dispersive, and Purcell enhancements to $T_1$ and $T_2$ shown in Figs. \ref{Q-Results-Waveguide}-\ref{Q-Results}.  

We have shown that the negotiation of a system's competing length-scales defines a unique fingerprint for defect physics. Such a fingerprint can serve as a powerful characterization tool, and can be used to test the foundations of glass physics. For example, we have demonstrated that defect decay and dephasing, observable using $\pi$-pulse and phonon echo \cite{Golding76,Enss96a,Enss96b}, directly probe the phonon DOS (e.g. Eq. \eqref{T1-ave} \& Fig. \ref{T1-cyl}) and reveal the nature of defect-defect interactions. Defect-induced electromagnetic noise reveal information about the DDOS, the system dimension, and fundamental origins of noise in qubits, and measurements of dissipation can probe system dimensionality, the phonon DOS, and energy dependence of the DDOS. A collection of such measurements can isolate and determine each of the parameters entering the standard tunneling state model. Thus, the TSM and its alternatives \cite{Yu88,Yu89,Leggett91,Vural11,Leggett13}, which give contrasting predictions in reduced dimensional systems \cite{Fu89}, can be put to the test.  

In closing, we have demonstrated that an ever-present source of noise and dissipation, engendered by low-energy defect centers, hinges sensitively on system scale and geometry. Our results show that this noise and dissipation can be reduced in mesoscale systems, suggesting that thoughtful mode engineering may enable unprecedented levels of performance in an array of cutting-edge technologies.  

\acknowledgements
R. B. and P. R. would like to thank William Renninger, Prashanta Kharel, Shai Gertler, Eric Kittlaus, Michel Devoret, Rob Schoelkopf, and Yiwen Chu for a number stimulating discussions and thoughtful suggestions. Primary support for this work was provided by NSF MRSEC DMR-1119826. The authors also acknowledge the support of Yale University start-up funding. 
F. I. acknowledges financial support from the EU through the Career Integration Grant (CIG) No. 631571 and support by the DFG through the DIP program (FO 703/2-1).
\begin{widetext}
\appendix 
\section{Derivation of $T_1^{-1}$}
\label{T1-Appendix}
In this section we derive the upper state lifetime for a defect coupled to a system's acoustic field. We begin our derivation by computing the transition amplitude for the coupled phonon-defect system to go from an initial state $| i \rangle = |e\rangle \otimes | \Psi_i \rangle$ at time $t_i$ to a final state $| f \rangle = |g\rangle \otimes | \Psi_f \rangle$ at time $t_f$ where the states $| \Psi_i \rangle$ and $| \Psi_f \rangle$ are energy eigenstates of the uncoupled phonon system. Formally, this amplitude can be written as 

\begin{equation}
\label{ }
c_{i \to f} = \langle f | U_I(t_f,t_i) |i \rangle
\end{equation}
where $U_I(t_f,t_i)$ is the time evolution operator in the interaction picture. The time evolution operator can be written as $U_I(t_f,t_i) = \mathcal{T} \exp\{- \frac{i}{\hbar} \int_{t_i}^{t_f} dt \ H^I_{\rm int}(t) \}$ where $H^I_{\rm int}$ is the interaction Hamiltonian in the interaction picture. In the weak coupling approximation the transition amplitude takes the form

\begin{equation}
\label{ }
c_{i \to f} \approx - \frac{i}{\hbar} \int_{t_i}^{t_f} dt \langle f | H^I_{\rm int}(t) |i \rangle. 
\end{equation}
The probability of deexcitation, for the process described above, is given by the modulus square of the transition amplitude. The total probability of deexcitation, via emission into all channels, is given by averaging over the initial state of the phonons and summing over all final states

\begin{equation}
\label{ }
P^{tot}_{e \to g} \approx \frac{1}{\hbar^2} \sum_{\Psi_f} \sum_{\Psi_i} p_i 
 \int_{t_i}^{t_f} dt  \int_{t_i}^{t_f} dt' 
 \langle g| \otimes \langle \Psi_f | H^I_{\rm int}(t) |e \rangle \otimes | \Psi_i \rangle
 \langle e| \otimes \langle \Psi_i | H^I_{\rm int}(t') |g \rangle \otimes | \Psi_f \rangle. 
\end{equation}
Since $\sum_{\Psi_i} p_i  | \Psi_i \rangle \langle \Psi_i | = \hat{\rho}$ is the initial density matrix of the phonon field, $\sum_{\Psi_f} | \Psi_f \rangle \langle \Psi_f | = {\bf I}$, and  
the relevant component of the interaction Hamiltonian is proportional to $\sigma_x$ and the strain field the transition probability is given by 
\begin{equation}
\label{ }
P^{tot}_{e \to g} \approx \frac{1}{\hbar^2} \frac{\Delta_0^2}{E^2} 
 \int_{t_i}^{t_f} dt  \int_{t_i}^{t_f} dt' e^{i E/\hbar(t-t')}
  \langle \boldsymbol{\gamma}:\underline{\xi}(t',{\bf r}) \boldsymbol{\gamma}:\underline{\xi}(t,{\bf r})   \rangle 
\end{equation}
where $\langle g | \sigma^I_x(t) | e \rangle = e^{-i E t/\hbar}$ has been used,  $\langle ...\rangle \equiv {\rm tr}  \{ \hat{\rho} ... \}$, $\underline{\xi}(t',{\bf r})$ is the freely evolving strain field, and ${\bf r}$ is the position of the defect.

We define the strain correlation function $G^+(t,t') \equiv \langle \boldsymbol{\gamma}:\underline{\xi}(t,{\bf r}) \boldsymbol{\gamma}:\underline{\xi}(t',{\bf r}) \rangle$ that only depends upon the difference in time arguments in steady-state. When $t_f,-t_i \to \infty$ a change of variables gives the decay rate $T_1^{-1}$, i.e. $(P^{tot}_{e \to g}+P^{tot}_{g \to e})/(t_f-t_i)$, as 
\begin{equation}
\label{ }
T_1^{-1} \approx \frac{1}{\hbar^2} \frac{\Delta_0^2}{E^2} 
(G^+(E/\hbar) + G^+(-E/\hbar)).
\end{equation}
For a phonon bath in thermal equilibrium the fluctuation-dissipation relation can be applied to reduce the expression above to
\begin{equation}
\label{ }
T_1^{-1} \approx \frac{2}{\hbar} \frac{\Delta_0^2}{E^2} \coth\frac{E}{2 k_B T}
{\rm Im}G(E/\hbar)
\end{equation}
 where $G(\omega)$ is the retarded Green's function related to $\boldsymbol{\gamma}:\underline{\xi}(t,{\bf r}_j)$. $G(t,t')$ can be derived from the Green's function for the displacement field $g_{lm}(x,x')$

\begin{equation}
\label{ }
[(\rho \partial_t^2 + \rho \Gamma \partial_t)\delta^l_i -  \partial_{j} C_{i}^{\ jkl} \partial_{k}]g_{lm}(x,x') = \delta^4(x-x') \delta_{im}
\end{equation}
where $C^{ijkl}$ is the system's elastic tensor, and where we've assumed that the phonons experience a linear dissipation $\Gamma$. 

We write $g_{lm}$ as a Fourier transform $ \tilde{g}_{lm}(\omega,{\bf x},{\bf x}') = (1/2\pi) \int d \omega e^{-i \omega t} {g}_{lm}(t,{\bf x};0,{\bf x}')$. The spatial dependence can be obtained by decomposing ${\bf g}$ into normal modes 

\begin{equation}
\label{ }
\tilde{\bf g}(\omega,{\bf x},{\bf x}') = \sum_{\bf q} A_{\bf q} \bf{u}_{\bf q}({\bf x}).
\end{equation}
Plugging this expansion into the equation above, and using the eigenvalue and orthonormality properties of the eigenfunctions, results in an expression for $A_{\bf q}$ 
yielding the following representation for $\tilde{\bf g}$
\begin{equation}
\label{ }
\tilde{\bf g}(\omega,{\bf x},{\bf x}') = \sum_{\bf q} \frac{\bf{u}_{\bf q}({\bf x}) \bf{u}^*_{\bf q}({\bf x}')}{\Omega_{\bf q}^2 -  i \omega \Gamma_{\bf q} -\omega^2}.
\end{equation}
Contracting each vector eigenfunction ${u}^k_{\bf q}$ with $\gamma_{ik}\partial_i$ gives $G$
\begin{equation}
\label{ }
G(\omega,{\bf x},{\bf x}') = \sum_{\bf q} \frac{\gamma: \underline{\xi}_{\bf q}({\bf x}) \gamma: \underline{\xi}_{\bf q}^*({\bf x}') }{\Omega_{\bf q}^2 -  i \omega \Gamma_{\bf q} -\omega^2} 
\end{equation}
which has an imaginary part at coincidence given by 

\begin{equation}
\label{ }
{\rm Im} G(\omega,{\bf x},{\bf x}) = \sum_{\bf q}  \frac{ \omega \Gamma_{\bf q} |\gamma: \underline{\xi}_{\bf q}({\bf x})|^2 }{(\Omega_{\bf q}^2 -\omega^2)^2 + (\omega \Gamma_{\bf q})^2}.
\end{equation}
The equation above leads to the decay rate for the defect given by 

\begin{equation}
\label{T1-formal}
T_1^{-1}(E) \approx \frac{2}{\hbar}  \sum_{\bf q} \frac{\Delta_0^2}{E^2} \coth\frac{E}{2 k_B T} \frac{ \Gamma_{\bf q} (E/\hbar) |\gamma: \underline{\xi}_{\bf q}({\bf r})|^2 }{(\Omega_{\bf q}^2 - (E/\hbar)^2)^2 + ((E/\hbar) \Gamma_{\bf q})^2} 
\end{equation}
which agrees with Eq. \eqref{T1} in the limit $\Gamma_{\bf q} \to 0$ and with Eq. \eqref{T1,cav} when averaged over defect orientations and positions.

\subsection{Waveguides}
In this section we compute the decay rate for a defect in a 2$D$ waveguide. We begin from Eq. \eqref{T1-ave}. The mode index ${\bf q}$ can be represented as $\{m,{\bf k}\}$ where $m$ is an index labeling the eigenfunctions describing the elastic field in the dimension normal to the plane, and ${\bf k}$ is a wavevector in the plane. Hence, the mode sum $\sum_{\bf q}$ is given by $\sum_m (A/4\pi^2)\int d^2 k$ where $A$ is the area of the plane, to be taken to infinity at the end of the calculation, and the mode eigenfrequencies can represented as $\Omega_{\bf q} \equiv \Omega_m({\bf k})$. Using the delta function identity listed inline above Eq. \eqref{T1-guided wave-1D} we find 
\begin{align}
\label{}
\langle T_1^{-1}(E) \rangle_V = & \frac{1}{L} \sum_{m,j,\eta} \int d^2 k \frac{\Delta_0^2}{2 \hbar \rho}  \frac{\gamma^2_\eta}{v_\eta^2} e_{mj,\eta} \coth \left(\frac{E}{2 k_B T}\right)
\frac{\delta(k-|{\bf k}_{mj}|)}{|\hbar v^{mj}_g|}
\end{align}
where $|{\bf k}_{mj}|$ is defined inline above Eq. \eqref{T1-guided wave-2D}, and $mj$ is short for $m,{\bf k}_{mj}$. Eq. \eqref{T1-guided wave-2D} is obtained by evaluating the ${\bf k}$-integral and using $v^{mj}_p = \Omega_{m}(|{\bf k}_{mj}|)/|{\bf k}_{mj}|$. 

\subsection{Bulk medium with dissipation}
In Eq. \eqref{T1-formal} we have derived the formal expression for the decay rate of a defect interacting with a collection of lossy phonon modes. It is interesting to evaluate the expression for $T_{1,{\rm min}}^{-1}$ in the infinite volume limit where the sum over modes becomes an integral. The result for $T_1$ in this limit depends upon the physical origin of the phonon decay. If the phonon dissipation is assumed to arise from local absorption due to the intrinsic losses present in the material we find the decay rate 
\begin{equation}
\label{ }
T_1^{-1}(E) \approx \frac{2}{2\pi^2\hbar \rho}  \sum_\eta \frac{\gamma^2_\eta}{v_\eta^5} \coth\frac{E}{2 k_B T} \int_0^\Lambda d\Omega  \frac{ \Gamma (E/\hbar) \Omega^4}{(\Omega^2 - (E/\hbar)^2)^2 + ((E/\hbar) \Gamma)^2}
\end{equation}
where we have introduced a high-frequency cutoff $\Lambda$ representing the defect `size'. If we take $\Gamma$ to be constant and assume $\Lambda \gg E/\hbar,\Gamma$ we find 
\begin{equation}
\label{T1-dissipative}
T_1^{-1}(E) \approx \frac{1}{\pi^2\hbar \rho}  \sum_\eta \frac{\gamma^2_\eta}{v_\eta^5}\frac{E}{\hbar} \coth\frac{E}{2 k_B T} \bigg[\Lambda \Gamma + \frac{\pi \Gamma}{4} \frac{{\rm Re}[\tilde{\omega}^3]}{{\rm Im}[\tilde{\omega}]{\rm Re}[\tilde{\omega}]} + O(\Lambda^{-1}) \bigg]
\end{equation}
where $\tilde{\omega} = \sqrt{(E/\hbar)^2 + i \Gamma(E/\hbar)}$. 

Notice that the decay rate is composed of a potentially large cutoff-dependent term when the acoustic medium is assumed to be lossy, and a cutoff independent term. This cutoff dependent contribution to the decay rate is well-known in the study of spontaneous emission of atoms embedded in absorbing dielectrics where it is attributed to non-radiative decay through near-field interactions \cite{Scheel99}. This interpretation is consistent in the acoustic case treated here as the cutoff-dependent part of the decay rate arises entirely from the $E = 0$ component of $T^{-1}_1$, i.e. from static elastic fields, the elastic equivalent of the electrostatic dipole field. Systems with large sources of intrinsic acoustic dissipation may require a more general treatment than that leading to Eq. \eqref{T1}. As an example consider phonon-phonon scattering in glass which leads to decay rates of order $(2\pi) 1$ Hz \cite{Goryachev13} at 3.8 K for GHz frequency phonons. For $\Lambda = 0.1$ nm the cutoff-dependent contribution to $T_1^{-1}$ is four orders of magnitude smaller than the cutoff-independent component, and hence Eq. \eqref{T1} gives a quantitative estimation of the defect decay rate in a dissipative bulk. When phonon dissipation arises from defects (all remaining defects) we find that GHz phonons and for $\Lambda = 0.1$ nm that the cutoff-dependent term is comparable to Eq. \eqref{T1}. We plan to investigate this effect further in future work.

Dissipation of phonon modes occurs even in lossless media when energy leaks from a resonator into a supporting structure. For this case a decay rate can be added to the equation of motion for the phonon field to model the energy leakage from a mode. For such a system the decay rate will scale as $\Gamma \sim -\frac{4 v}{L} \ln r$ where $v$ is the sound speed, $L$ is the characteristic size of the resonator, and $r$ is the reflection coefficient, representing the fraction of energy retained in the system for each cycle. Unlike the previous example, the dissipation is not distributed throughout the resonator, i.e. the losses occur as energy leaks away upon reflection at the resonator-support interface. Hence, for this system the cutoff-dependent component of the decay rate is an artifact of the way we have modeled the cavity losses and thus should be subtracted. Indeed, for the case of an atom in an inhomogeneous dielectric a calculation of the decay rate is cutoff-independent so long as the dielectric imediately surrounding the atom is not lossy \cite{Tomas01}. Also, note that the decay rate vanishes in the infinite volume limit, and hence Eq. \eqref{T1-dissipative} appropriately reduces to Eq. \eqref{T1} for a lossless medium in the infinite volume limit. 

\section{Elastic energy and angular averages of $\boldsymbol{\gamma}:{\underline{\xi}_{\bf q}}$ }
\label{Identity-Appendix}
The elastic energy of a system occupying volume $V$ is given by

\begin{align}
\label{ }
\mathcal{E} & = \mathcal{K}+\mathcal{V} \nonumber  = \frac{1}{2} \int_V d^3x \ [ \rho \dot{\bf u}^2 + C^{ijkl}\xi_{ij}\xi_{kl} ]
\end{align}
where $\mathcal{K}$ is the kinetic energy, $\mathcal{V}$ is the potential energy, and $C^{ijkl}$ is the elastic tensor. By decomposing the elastic field into   normal modes, using the orthonormality relation for the displacement field, integrating the potential energy term by parts, and using the eigenvalue equation for the normal modes one finds the following identities

\begin{equation}
\label{ }
\mathcal{K} = \mathcal{V} = \frac{1}{4} \sum_{\bf q} \hbar \Omega_{\bf q} (b_{\bf q} b^\dag_{\bf q} + b^\dag_{\bf q} b_{\bf q} ).
\end{equation}
For an isotropic medium the elastic tensor is $C^{ijkl} = \lambda \delta^{ij}\delta^{kl} + \mu(\delta^{ik} \delta^{jl} + \delta^{il} \delta^{jk})$ where $\lambda = \rho(v_\ell^2-2v_t^2)$ and $\mu = \rho v_t^2$. With a mode decomposition of the strain, the identity $\mathcal{E} = 2 \mathcal{V}$, and the orthonormality relation we find 

\begin{equation}
\label{Energy}
\mathcal{E} = \sum_{\bf q}\frac{1}{2} \hbar \Omega_{\bf q} (b_{\bf q} b^\dag_{\bf q} + b^\dag_{\bf q} b_{\bf q} ) \left[ \frac{1}{\Omega_{\bf q}^2} \int_V d^3x \  \rho ( v_\ell^2 |{\rm tr} \underline{\xi}_{\bf q}|^2 + 2 v_t^2 (\underline{\xi}_{\bf q}:\underline{\xi}_{\bf q}^*-|{\rm tr} \underline{\xi}_{\bf q}|^2 ) \right].
\end{equation}
The quantity in square brackets is equal to 1, and which we interpret as the sum of the fractions of elastic energy in compressional $e_{{\bf q}\ell}$ and shear $e_{{\bf q}t}$ motion of the ${\bf q}$th mode
\begin{align}
\label{Fractions-a}
e_{{\bf q}\ell} & =  \frac{1}{\Omega_{\bf q}^2} \int_V d^3x \  \rho v_\ell^2 |{\rm tr} \underline{\xi}_{\bf q}|^2 \\
\label{Fractions-b}
e_{{\bf q}t} & = \frac{2}{\Omega_{\bf q}^2} \int_V d^3x \  \rho v_t^2 (\underline{\xi}_{\bf q}:\underline{\xi}_{\bf q}^*-|{\rm tr} \underline{\xi}_{\bf q}|^2). 
\end{align}

Now we evaluate $\langle |\boldsymbol{\gamma}:{\underline{\xi}_{\bf q}}|^2 \rangle_V$. Using $ \boldsymbol{\gamma}: \underline{\xi}
= \tilde{\gamma} [(1-2\zeta) {\rm tr} \underline{\xi} + 2 \zeta \hat{n} \cdot \underline{\xi} \cdot \hat{n} ]$ \cite{Anghel07} the $\hat{n}$-orientation average of $ |\boldsymbol{\gamma}:{\underline{\xi}_{\bf q}}|^2 $ can be performed giving

\begin{equation}
\label{ }
\int \frac{d\varphi}{4\pi} |\boldsymbol{\gamma}:{\underline{\xi}_{\bf q}}|^2 
=
 \gamma_\ell^2 |{\rm tr} \underline{\xi}_{\bf q}|^2 + 2 \gamma_t^2 (\underline{\xi}_{\bf q}:\underline{\xi}_{\bf q}^*-|{\rm tr} \underline{\xi}_{\bf q}|^2 ). 
\end{equation}
Finally, averaging the result above over the system volume we arrive at Eq. \eqref{Ident-1} if we assume that the density throughout the system is uniform.  

\section{Small Mode Volume Limit}
\label{Rabi-Appendix}
The analysis of defect-induced dissipation in the main body of the text applies to the scenario where a large number of defects interact with each phonon mode. However, for resonator systems as the system volume becomes small a point is reached where the acoustic modes can interact with a single defect, occurring when $N_{\rm res} \leqslant 1$. In this case the defects no longer act as a spin bath which can irreversibly absorb acoustic energy, and the defect-phonon system will undergo Rabi oscillation. 
For silica based systems 
\begin{equation}
\label{ }
N_{\rm res} \sim P \hbar \Gamma_{\bf q} V =  \frac{\Gamma_{\bf q}}{2\pi {\rm MHz}}\frac{V}{(1 \mu {\rm m})^3},
\end{equation}
meaning that the small mode volume limit can be achieved with 1 MHz linewidth modes in systems with volumes less than a $1 \ \mu$m$^3$. For the formula above the acoustic mode decay rate is determined by all other sources of loss in the system such as phonon-phonon scattering. 

The energy of systems with $N_{\rm res} = 1$ will oscillate between the defect and phononic degrees of freedom with the frequency

\begin{equation}
\label{ }
\hat{\Omega}_{\rm Rabi}^2 =  \frac{2}{\hbar \Omega_{\bf q}} \frac{\Delta_0^2}{E^2} |{\gamma}:{\underline{\xi}_{\bf q}}({\bf r})|^2 \left(\hat{N} + \frac{1}{2}\right) + (\Omega_{\bf q} - E/\hbar)^2
\end{equation}   
which is calculated from the Heisenberg equations (Appendix A) of the coupled system in the rotating wave approximation (RWA) \cite{EberlyTLABook}. The operator $\hat{N} = b_{\bf q}^\dag b_{\bf q} + \frac{1}{2} \sigma_z$ is conserved in the RWA. 

\section{Dipole-dipole correlation function}
\label{Correlation-Appendix}
In this section we compute the dipole-dipole correlation function for the defects using the QRT. 
For weak coupling the dipole-dipole correlation function separates into two terms
\begin{align}
\label{ }
\langle  \boldsymbol{\delta \mathcal{P}}_i(t) \boldsymbol{\delta\mathcal{P}}_j(t') \rangle = & \delta_{ij} {\bf d} {\bf d} \bigg[  \frac{\Delta_{0}^2}{E^2} \langle \sigma_x(t) \sigma_x(t') \rangle \nonumber \\
& +\frac{\Delta^2}{E^2} ( \langle  \sigma_z(t)  \sigma_z(t')\rangle -w_0^2(E))   \bigg]
\end{align}
where we've suppressed the defect labels $i$ and $j$ on defect parameters, and ${w_{0}(E) \equiv - \tanh \frac{E}{2 k_B T}}$ is the thermal equilibrium value of  $\sigma_z$.  The correlation functions above can be approximated using the quantum regression theorem (QRT) \cite{MandelWolf}, positing that the two-time correlation function of an operator $A$ $\langle A(t) A(0) \rangle$ satisfies the same equation of motion as the mean value $\langle A(t) \rangle$, giving
\begin{align}
\label{SigX}
 \langle \sigma_x(t) \sigma_x(t') \rangle \approx & \frac{1}{2}(1+w_0(E))e^{i\frac{E}{\hbar}(t-t') - \frac{|t-t'|}{T_2} }  \\
 & + \frac{1}{2}(1-w_0(E))e^{-i\frac{E}{\hbar}(t-t') - \frac{|t-t'|}{T_2} } \nonumber 
 \\
 \label{SigZ}
  \langle  \sigma_z(t)  \sigma_z(t')\rangle \approx & e^{- \frac{|t-t'|}{T_1}} + w_0^2(E) 
  \left(1- e^{- \frac{|t-t'|}{T_1}} \right).
\end{align}
The correlation functions above satisfy the Pauli operator algebra at equal times, and at large time separations, i.e. for $|t-t'| \gg T_1$ (or $T_2)$ the two operators appearing in $\langle  \sigma_z(t)  \sigma_z(t')\rangle$ are completely uncorrelated and hence the correlation function factorizes $\langle  \sigma_z(t)  \sigma_z(t')\rangle = \langle  \sigma_z(t) \rangle \langle  \sigma_z(t')\rangle = w_0^2(E)$. The QRT is valid in the weak coupling limit. 

\section{Derivation of resonant absorption}
\label{Resonant-Appendix}
In this section we derive resonant absorption of a driven acoustic mode. 
\subsection{Heisenberg equations of motion}
To orient the reader and establish notation we first give the Heisenberg equations of motion for the coupled defect-phonon system in full generality
\begin{align}
\label{}
    \dot{\sigma}_{z,j}   &   =  \frac{2}{\hbar}\sum_{\bf q} (\gz b_{\bf q} + \gz^* b^\dag_{\bf q}) \sigma_{y,j} 
    \\
   \dot{\sigma}_{y,j}   &   = \frac{1}{\hbar} [E'_j +  2 \sum_{\bf q}(\g b_{\bf q} + \g^* b^\dag_{\bf q})]\sigma_{x,j} - \frac{2}{\hbar} 
   \sum_{\bf q}(\gz b_{\bf q} + \gz^* b^\dag_{\bf q}) \sigma_{z,j} \\
   \dot{\sigma}_{x,j}   &   = -\frac{1}{\hbar} [E'_j + 2 \sum_{\bf q}(\g b_{\bf q} + \g^* b^\dag_{\bf q})]\sigma_{y,j} \\
   \dot{b}_{\bf q}       &   = - i \Omega_{\bf q} b_{\bf q} - 
   \frac{i}{\hbar} \sum_j [ \g^* \sigma_{z,j} + \gz^* \sigma_{x,j}]
 \end{align} 
where the shorthand  
$\gz \equiv \frac{\Delta_{0j}}{E_j} \sqrt{\frac{\hbar}{2 \Omega_{\bf q}}}
 \boldsymbol{\gamma}:{\underline{\xi}_{\bf q}}({\bf r}_j)$
 and $\g \equiv \frac{\Delta_{j}}{E_j} \sqrt{\frac{\hbar}{2 \Omega_{\bf q}}}
 \boldsymbol{\gamma}:{\underline{\xi}_{\bf q}}({\bf r}_j)$
 has been introduced. The equations for the coupled defect-photon system can be derived by taking $\gz = -i \frac{\Delta_{0j}}{E_j} \sqrt{\frac{\hbar \Omega_{\bf q}}{2 }}
 {\bf d}\cdot {\bf E}_{\bf q}({\bf r}_j)$
 and ${\g = -i\frac{\Delta_{j}}{E_j} \sqrt{\frac{\hbar \Omega_{\bf q}}{2}}
  {\bf d}\cdot {\bf E}_{\bf q}({\bf r}_j)}$ and $b_{\bf q} \to a_{\bf q}$, the annihilation operator for the EM field, in the equations above. 
\subsection{Bloch equations}
For the purposes of deriving resonant absorption we work with the Bloch equations describing the mean-field dynamics of the interaction of a single phonon mode with the defect ensemble. In addition, we neglect the diagonal coupling $\Delta$ which plays a minor role in this process. The Bloch equations can be derived from the Heisenberg equations of motion in leading order perturbation theory resulting in
\begin{align}
\label{}
  \dot{S}_z  & = -\frac{1}{T_1} (S_z-w_0) 
  -\frac{i 2}{\hbar} (\gzn \beta_{\bf q} S^+ - \gzn^* \beta^\dag_{\bf q} S^-) \\
  \dot{S}^+  &  = \left(i E/\hbar -T^{-1}_2 \right) S^+ - \frac{i}{\hbar} \gzn^* \beta^\dag_{\bf q} S_z \\
  \dot{\beta}_{\bf q} & = ( - i \Omega_{\bf q} - \Gamma_{\bf q}/2) \beta_{\bf q} - \frac{i}{\hbar} \gzn S^- + f_{\bf q}
\end{align}
where $S_k \equiv \langle \sigma_k \rangle$,  $S^\pm = (S_x \pm i S_y)/2$, the effect of thermal fluctuations of the phonon field have been accounted for in the decay rates $T_1$ and $T_2$, $\beta_{\bf q}$ is the driven component of the phonon field, $f_{\bf q} \propto e^{-i \omega t}$ is an external drive, and the RWA has been used which is valid so long as $\Omega_{\bf q} \gg \Gamma_{\bf q}$. The dephasing time $T_2$ results from thermalization by the phonon field and by other thermally active defects in the system. The latter is unimportant when $P k_B T V < 1$, i.e. there are no thermally active defects present in the system, which is feasible with very small mode volumes and low temperatures. For silica $P k_B T V \approx 75$ for $T=10$ mK and $V = 1\ \mu$m$^3$.  

With a strong external drive oscillating at $\omega$ we look for solutions with $\beta_{\bf q}$ and $S^-$ 
 both oscillating as $e^{- i \omega t}$ and $S_z$ time-independent. In this approximation, the solution for $S^+$ and $S_z$ are given by 

\begin{align}
\label{}
  S_z  & = \frac{w_0}{1 +  \frac{4 |\gzn|^2 T_1 T_2}{\hbar^2} \frac{1}{1+(E/\hbar - \omega)^2T_2^2} |\beta_{\bf q}|^2 }  \\
 S^+ & = - \frac{i}{\hbar} \gzn^* \frac{T_2}{ 1-iT_2(E/\hbar-\omega)} \beta^\dag_{\bf q} S_z.
\end{align}    
When plugged into the equation of motion for the phonon, these solutions account for the back reaction of the defect on the phonon mode. This is manifested as a frequency shift and a dissipation function: $[-i \Delta \Omega^{\rm res}_{\bf q} - \Gamma^{\rm res}_{\bf q}/2] \beta_{\bf q}$

\begin{align}
\label{}
  - i \omega \beta_{\bf q}  & = [- i (\Omega_{\bf q} + \Delta \Omega^{\rm res}_{\bf q}) - (\Gamma_{\bf q} + \Gamma^{\rm res}_{\bf q})/2 ]\beta_{\bf q} + f_{\bf q}  
\end{align} 

\begin{align}
\label{resAbs-1defect}
\Gamma^{\rm res}_{\bf q} & = - 2 \frac{|\gzn|^2}{\hbar^2}\frac{T_2}{1+ T_2^2(\omega- E/\hbar)^2} S_z \\
									& =  2 \frac{|\gzn|^2}{\hbar^2}\frac{\tanh\left( \frac{E}{2 k_B T} \right)}{\sqrt{1+\frac{4 |\gzn|^2 T_1 T_2}{\hbar^2} |\beta_{\bf q}|^2}} \frac{\tilde{T}_2}{1+ \tilde{T}_2^2(\omega- E/\hbar)^2}
\end{align} 
where $\tilde{T}_2 = \sqrt{1+\frac{4 |\gzn|^2 T_1 T_2}{\hbar^2} |\beta_{\bf q}|^2} T_2$. Eq. \eqref{resAbs-1defect} gives the dissipation rate for resonant absorption of acoustic energy by a single defect. In the case when the energy of a large number of defects fall within $1/2\pi T_2$ of the mode frequency the dissipation rate can be calculated by taking the ensemble average of Eq. \eqref{resAbs-1defect} with respect to all defect properties. Such an averaging, and using $\Omega_{\bf q} \gg \Gamma_{\bf q}$, results in Eq. \eqref{resAbs-Gen}. 

\section{Derivation of relaxation absorption}
\label{Relaxation-Appendix}
In this section we derive the phonon dissipation rate due to relaxation absorption. 
We begin with the observation, with some rearrangement of Eq. \eqref{Hamiltonian}, that the energy level of a given defect is modulated by an incident strain field 
as $E_j \to E_j + 2 \frac{\Delta_j}{E_j} \boldsymbol{\gamma}_j: \underline{\xi}({\bf r}_j)$.
This energy-level modulation will drive the level inversion which can be accounted for by Taylor expanding $w_0(E)$ for small strain in the Bloch equation for the $S_z$
\begin{equation}
\label{Sz-RelAbs}
 \dot{S}_z   \approx -\frac{1}{T_1} \left( S_z-w_0-  2 \frac{\partial w_0}{\partial E}  \frac{\Delta}{E} \boldsymbol{\gamma}: \underline{\xi}({\bf r}) \right) 
\end{equation}
where the off-diagonal coupling, proportional to $\gz$,  and the suffix $j$ has been dropped. 

The strain field on the right hand side of Eq. \eqref{Sz-RelAbs} drives oscillations of the level inversion, and in turn, level inversion oscillations lead to the radiation of phonons in a random direction. This acoustic radiation can be computed by finding the solution of $S_z$ and plugging it's solution into the equation of motion for the mean acoustic field

\begin{equation}
\label{ }
 \dot{\beta}_{\bf q}  = (- i \Omega_{\bf q} - \Gamma_{\bf q}/2) \beta_{\bf q} - \frac{i}{\hbar} g_{\bf q} S_z 
\end{equation}  
where the diagonal coupling has been reintroduced and $\gz$ has been set to zero. The oscillating component of the level inversion $\delta S_z$
is given by 

\begin{equation}
\label{ }
\delta S_z \approx \frac{2}{1-i\Omega_{\bf q} T_1} \frac{\partial w_0}{\partial E} g_{\bf q} \beta_{\bf q}. 
\end{equation}
Plugging this expression into the equation of motion for the phonon, i.e. using the RWA, results in dissipation and a frequency shift
\begin{align}
\label{ }
- \frac{i}{\hbar} g_{\bf q} \delta S_z = & (-i\Delta \Omega^{\rm rel}_{\bf q} -\Gamma^{\rm rel}_{\bf q}/2) b_{\bf q} 
\end{align}
where
\begin{align}
\label{ }
\Delta \Omega^{\rm rel}_{\bf q}  & = \frac{2}{\hbar} |g_{\bf q} |^2  \frac{1}{1+\Omega_{\bf q}^2 T_1^2}  \frac{\partial w_0}{\partial E}
\\
\label{relAbs-1defect}
\Gamma^{\rm rel}_{\bf q} & = - \frac{4}{\hbar} |g_{\bf q} |^2  \frac{\Omega_{\bf q} T_1}{1+\Omega_{\bf q}^2 T_1^2}  \frac{\partial w_0}{\partial E} 
\end{align}
which is the frequency shift and dissipation from relaxation absorption for a single defect. When averaged over all defect properties Eq. \eqref{relAbs-1defect} reduces to Eq. \eqref{relAbs} in the main text.

$\Gamma^{\rm rel}_{\bf q}$ can be interpreted as the energy lost from the phonon mode as it drives a defect, and the defect reradiates that energy into a random direction with a strength set by $T^{-1}_1$. In the high frequency limit notice that the defect decay is proportional to $T_1^{-1}$. Hence, for a cavity based system only resonant defects will appreciably reradiate the phonon's energy. Given the exponential suppression of $\Gamma^{\rm rel}_{\bf q}$ with $E$ by $\frac{\partial w_0}{\partial E}  = -\frac{1}{2 k_B T} \sech^2 \frac{E}{2 k_B T}$, the dominant contribution to relaxation absorption in a cavity based system is given by the defects that are resonant with the fundamental mode $\Omega_0$. Hence, the quality factor is given by

\begin{equation}
\label{resAbs-Cav1}
\frac{1}{Q_{{\rm rel},{\bf q}}} \approx \frac{4}{\hbar} |g_{\bf q} |^2  \frac{1}{\Omega_{\bf q}^2 T_1}  \frac{1}{2 k_B T} \sech^2
\frac{\hbar \Omega_1}{2 k_B T} 
\end{equation} 
with $T_1$ evaluated at $E = \hbar \Omega_1$, and hence relaxation absorption is exponentially suppressed for low temperatures where $\hbar \Omega_1/k_B T \gg 1$. After averaging over defect properties it can be shown that Eq. \eqref{resAbs-Cav1} is the first term of Eq. \eqref{relAbs-Cavity-2}.
\end{widetext}


\begin{references}
\bibitem{Footnote-1}
The equality ${\gamma_\ell^2/v_\ell^2 = \gamma_t^2/v_t^2}$ is satisfied for the range of deformation potential values measured in silica.
\bibitem{Footnote-2}
The last equality holds when the orientation a defect's electric dipole and deformation potential are uncorrelated on average.
\end{references}
\end{document}